\documentclass{IEEEoj}

\usepackage{cite}
\usepackage{amsmath,amssymb,amsfonts}
\usepackage{algorithmic}
\usepackage{graphicx,color}
\usepackage{amssymb}
\usepackage{amsfonts}
\usepackage{amsmath}
\usepackage{algorithm}
\usepackage{algorithmic}
\usepackage{tikz}
\usepackage{graphicx,subfigure,cite,multicol,multirow,diagbox,booktabs,array}
\usepackage{color}
\usepackage{stfloats}
\usepackage{textcomp}
\def\BibTeX{{\rm B\kern-.05em{\sc i\kern-.025em b}\kern-.08em
		T\kern-.1667em\lower.7ex\hbox{E}\kern-.125emX}}
\AtBeginDocument{\definecolor{ojcolor}{cmyk}{0.93,0.59,0.15,0.02}}
\def\OJlogo{\vspace{-4pt}\hskip-4pt\includegraphics[height=18pt]{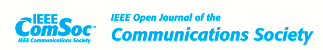}}
\begin{document}
	\receiveddate{XX Month, XXXX}
	\reviseddate{XX Month, XXXX}
	\accepteddate{XX Month, XXXX}
	\publisheddate{XX Month, XXXX}
	\currentdate{11 January, 2024}
	\doiinfo{OJCOMS.2024.011100}
	
	\title{Joint Power Allocation and Beamforming Design for Active IRS-Aided Directional Modulation Secure Systems \\ }
	\author{Yifan~Zhao\IEEEauthorrefmark{1} , ~Xiaoyu~Wang
		\IEEEauthorrefmark{2}, ~Kaibo~Zhou\IEEEauthorrefmark{2}, ~Xuehui~Wang  \IEEEauthorrefmark{1}, ~Yan~Wang \IEEEauthorrefmark{1},~Wei Gao\IEEEauthorrefmark{1},~Ruiqi~Liu\IEEEauthorrefmark{3} ~and~Feng~Shu \IEEEauthorrefmark{1,4}
	}
	\affil{School of Information and Communication
		Engineering, Hainan University, Haikou 570228, China}
	\affil{Tyr Systems Laboratory, China Academy of Information and Communications Technology, Beijing 100000, China}
	\affil{Wireless and Computing Research Institute, ZTE
		Corporation, Beijing 100029, China}
	\affil{School of Electronic and Optical Engineering, Nanjing University of Science and Technology, Nanjing 210094, China}
	\corresp{CORRESPONDING AUTHOR: Xuehui Wang, Wei Gao and Feng Shu (e-mail: wangxuehui0503@163.com, gaowei@hainanu.edu.cn and shufeng0101@163.com.)}
	\authornote{This work was supported in part by the National Key Research and Development Program of China under Grant 2023YFF0612900. }
	\markboth{Preparation of Papers for IEEE OPEN JOURNALS}{Author \textit{et al.}}

	\begin{abstract}
		Since the secrecy rate (SR) performance improvement obtained by secure directional modulation (DM) network is limited, an active intelligent reflective surface (IRS)-assisted DM network is considered to attain a high SR. To address the SR maximization problem, a novel method based on Lagrangian dual transform and closed-form fractional programming algorithm (LDT-CFFP) is proposed, where the solutions to base station (BS) beamforming vectors and IRS reflection coefficient matrix are achieved. However, the computational complexity of LDT-CFFP method is high . To reduce its complexity, a blocked IRS-assisted DM network is designed. To meet the requirements of the network performance, a power allocation (PA) strategy is proposed and adopted in the system. Specifically, the system power between BS and IRS, as well as the transmission power for confidential messages (CM) and artificial noise (AN) from the BS, are allocated separately.  Then we put forward null-space projection (NSP) method, maximum-ratio-reflecting (MRR) algorithm and PA strategy (NSP-MRR-PA) to solve the SR maximization problem. The CF solutions to BS beamforming vectors and IRS reflection coefficient matrix are respectively attained via NSP and MRR algorithms. For the PA factors, we take advantage of exhaustive search (ES) algorithm, particle swarm optimization (PSO) and simulated annealing (SA) algorithm to search for the solutions. From simulation results, it is verified that the LDT-CFFP method derives a higher SR gain over NSP-MRR-PA method. For NSP-MRR-PA method, the number of IRS units in each block possesses a significant SR performance. In addition, the application PA strategies, namely ES, PSO, SA methods outperforms the other PA strategies with fixed PA factors. 
	\end{abstract}
	
	\begin{IEEEkeywords}
		Directional modulation, active intelligent reflective surface, Lagrangian dual transformation, fractional programming, power allocation
	\end{IEEEkeywords}
	
	\maketitle
	
	\section{INTRODUCTION}
	In the age of information explosion, secure transmission of confidential message (CM) is becoming increasingly important. Due to the broadcast characteristic of wireless communication,  it is challenging to ensure that the CM can be accurately received by the excepted users. Directional modulation (DM) can improve the CM strength of expected legitimate users through beamforming design and reduce the interference of eavesdropping users on legitimate users via artificial noise (AN), so that DM has a capability to modify the security effectiveness of wireless communication\cite{TY9530396,WSM8315448}. 
	However, as the number of users surges in DM networks, there exists spectrum resource limitations and signal interference, which poses some serious challenges to the secrecy rate (SR) performance and reliability of conventional DM \cite{ST8839973,TYarc,QYL8290952,QJP9120206}.

	Because of low-power and low-cost characteristic, intelligent reflective surface (IRS) technology has garnered considerable interest to address these challenges \cite{EMA9086766,DLL9998527,CM8723525,YYH10284897,LRQ9679804}. By controlling the reflective phase and amplitude of units \cite{WXH9687669,LXH10190735,LRQ10298069,FL10036440,HM9740154,LRQ9955484}, IRS can significantly boost the rate performance of communication systems, so that the communication coverage is extended and power consumption is saved. The AN-aided secure multiple-input multiple-output (MIMO) wireless communication system was proposed in \cite{PCH9201173}, while the power limitation was taken into account. Besides, the block coordinate descent (BCD) algorithm was proposed to solve the problem of maximizing SR.  
	To enhance the robustness of an unmanned aerial vehicle wireless network aided by IRS, a robust SR maximization problem was formulated in \cite{YRL10288199}, and then an alternating optimization (AO) method was utilized to jointly design UAV transmit power, trajectory and IRS phase shifters.
	In addition, an IRS-aided secure simultaneous wireless information and power transfer system was considered in \cite{NHH9234391}. Based on an imprecise BCD method, penalty optimization minimization and complex manifold methods were put forward to jointly design the precoding matrix, the AN covariance and IRS phase shifters. 
	\cite{AH9399840} proposed an IRS-assisted spatial modulation system, which studied multi-direction beamforming with multiple antenna selection. To convey more information bits, an successive signal detection detector was designed to translate the received information.

	Considering the above advantages of IRS, IRS can be introduced to DM secure wireless network to create a more secure and reliable communication environment. So far, some related research has appeared. There was an IRS-assisted DM network designed in \cite{LLL9232092}, where a closed-form solution of SR was derived, and it was verified that the SR performance could get a bit gain by IRS. In \cite{LYQ10172219}, the authors made an investigation of an IRS-aided DM secure network with single-antenna users. Aiming at solving SR maximization problem, the authors designed CM beamforming vector via the Rayleigh ratio, and AN beamforming vector and phase shift matrix through generalized power iteration. Additionally, by only reflecting CM and no AN, a low computational complexity method with zero-forcing constraint was proposed to maximize receive power. Besides, an IRS-based DM network with multi-antenna users was considered in \cite{TY9530396}, where a transmit single conﬁdential bit stream  (CBS) from Alice to Bob could be added with the aid of IRS. To improve the SR, the authors respectively designed a high-performance general alternating iterative algorithm and a low-complexity null-space projection (NSP) algorithm to jointly optimize the transmit beamforming vectors and phase shift matrix.


	Obviously, from the above discussion about IRS-aided DM secure network, it is concluded that IRS can enhance the SR performance of DM secure wireless network. However, the above research focused on passive IRS. It is known that reflected signal experiences path fading twice, which is inevitable\cite{YG9658554,WXM8333706,LWG9963962}. To weaken the double-fading effect, active IRS appears, which can amplify reflected signal. Currently, active IRS has obtained extensive research \cite{KZY10235893}. An active IRS-aided multi-user wireless communication system was proposed in \cite{DLL9998527}, and for a maximum sum rate, a combination of fractional programming (FP) and AO was put forward to design the transmit beamforming at BS and active IRS reflective coefficient matrix. To further validate the amplification feature of active IRS, the authors in \cite{LJ10135161} considered an active IRS-assisted single-input single-output wireless network. While the closed-form solution to reflecting coefficient matrix was derived by maximum ratio reflecting (MRR) method and selective ratio reflecting (SRR) method. Compared with the equal-gain reflecting method in \cite{DLL9998527}, MRR and SRR methods could attain a higher rate gain. To boost the achievable rate of system proposed in \cite{DLL9998527}, the authors in \cite{LYQ10141981} designed three methods, i.e., maximizing the simplified signal-to-noise ratio plus Rayleigh-Ritz theorem (Max-SSNR-RR), generalized maximum ratio reflection (GMRR), maximizing SNR plus FP (Max-SNR-FP), to design the amplifying factors and the corresponding phases at active IRS. From the related simulation results, the proposed three methods perform better than MRR and SRR methods in \cite{DLL9998527}.

	Since the double-fading effect exists in cascaded channels related to passive IRS, the SR performance achieved by passive IRS-assisted DM secure network is limited, which is far from meeting the high SR requirement of current wireless secure network. Given that active IRS can bring significant gain by amplifying reflected signal to the wireless network, in order to improve SR performance of DM secure network aided by passive IRS, we consider that integrating active IRS into DM secure network forming an active IRS-assisted DM secure network in this paper. At present, the research on the combination network of active IRS and DM network is little, which is a topic worth exploring. 
	In addition, it is demonstrated that power allocation (PA) has a significant impact on SR performance\cite{CYW793310,ZJH10284767,KZ8661536,ZXK8051264}. Therefore, under the constraint of total power, we jointly design PA factors, transmit beamforming vectors, active IRS reflecting coefficient matrix to obtain maximum SR. The main contributions of our paper are summarized as follow: 
	

	\begin{enumerate}
		
		\item

		To obtain an excellent SR enhancement, a DM secure network assisted by active IRS is designed. We formulate SR maximization problem, which is subject to the total power constraint. Since transmit beamforming vectors for CM and AN, and active IRS reflecting coefficient matrix are coupled with each other, we take Lagrangian dual transform (LDT) to decouple the optimization variables. Afterwards, we take advantage of closed-form fractional programming (CFFP) algorithm to achieve the CF solutions to the variables. Simulation results show that LDT-CFFP method can harvest higher SR gain than the method in \cite{DLL9998527}, where the powers of BS and active IRS are subject to their respective power constraint. Besides, the computational complexity of LDT-CFFP method is $\mathcal{O}\{ I[\sqrt{KM}(2K^5M^5+K^4M^4)+\sqrt{N}(2N^4+N^3)+K^2M+KN+KM]\}$.

		\item

		To reduce the complexity of LDT-CFFP method, a blocked IRS-based DM secure network with PA strategy is proposed. Simultaneously, the PA of BS and IRS, CM and AN are considered. To solve the problem of maximizing SR, NSP is firstly applied to sub blocks, so that one block only reflect CM and no AN to Bob, and the other one only reflect AN to Eve. After that, the CF solutions to BS beamforming vectors are attained. MRR algorithm is adopted to attain the CF solutions to IRS reflection coefficient matrix. Meanwhile, three PA strategies called algorithm exhaustive search (ES), particle swarm optimization (PSO) and simulated annealing (SA) are utilized to search PA parameters. Simulation results present that in contrast to the other PA strategies with fixed PA factors, NSP-MRR-PA (ES), NSP-MRR-PA (PSO) and NSP-MRR-PA (SA) can achieve up to 27\% SR gain when the number of active IRS is 8. Furthermore, their computational complexity are  $\mathcal{O}\{NM+N+K_{ES}\}$, $\mathcal{O}\{NM+N+K_{PSO}I_{PSO}\}$,  $\mathcal{O}\{NM+N+K_{SA}I_{SA}\}$, which are lower than that of LDT-CFFP method.
	\end{enumerate}

	The remainder of the paper is arranged as follows. In Section II, we consider an active IRS-aided DM network and formulate the corresponding SR maximization problem. Section III 
	presents a high-performance LDT-CFFP method. Section IV shows a low-complexity NSP-MRR-PA method for blocked IRS-aided DM network. Section V shows simulation results, and Section VI presents the conclusion.

	$Notations$: In this paper, bold lowercase and uppercase
	letters represent vectors and matrices, respectively. Transpose and conjugate transpose operations are represented by $(\cdot)^T$ and $(\cdot)^H$, respectively. Euclidean paradigm, diagonal, expectation and real part operations are denoted as $\|\cdot\|$, diag$(\cdot)$, $E\{\cdot\}$ and $\mathfrak{R}\{\cdot\}$, respectively.

	\section{System Model}
	
	\begin{figure}[t]
		\centering
		\includegraphics[width=0.5\textwidth]{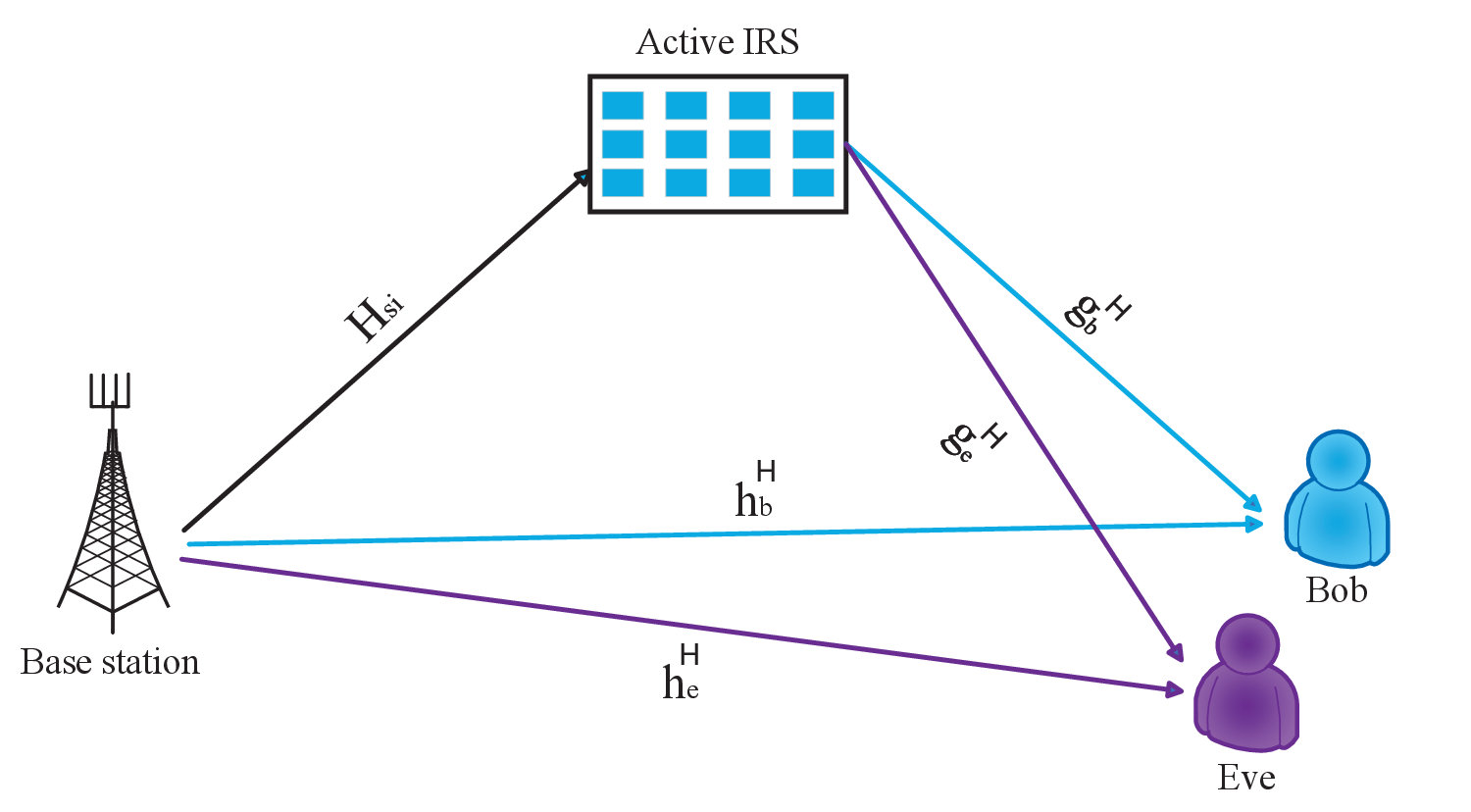}
		\caption{Active IRS-assisted DM secure communication system }
		\label{systemmodel}
	\end{figure}
	Fig.\ref{systemmodel} shows an active IRS-assisted DM system, which comprises of a BS with $M$ antennas, an active IRS with $N$ elements, a legitimate user (Bob) and an eavesdropper (Eve) with  single antenna, respectively. 
	$\mathbf{H}_{si}$ $\in$ $\mathbb{C}^{{N \times M}}$, $\mathbf{g}_b^H$ $\in$ $\mathbb{C}^{{1 \times N}}$, $\mathbf{g}_e^H$ $\in$ $\mathbb{C}^{{1 \times N}}$, $\mathbf{h}_b^H$ $\in$ $\mathbb{C}^{{1 \times M}}$ and $\mathbf{h}_e^H$ $\in$ $\mathbb{C}^{{1 \times M}}$ respectively denote the channels from BS to IRS, from IRS to Bob, from  IRS to Eve, from BS to Bob and from BS to Eve. 
	The transmit signal from BS is
	\begin{equation}
		\mathbf{x}=\mathbf{v}_bx_b+\mathbf{v}_ex_e,
	\end{equation}
	where $x_b$ and $x_e$ are the CM and AN, they satisfy $\mathbb{E}\left\{{|x_b|^2}\right\}=1$ and $\mathbb{E}\left\{{|x_e|^2}\right\}=1$. $\mathbf{v}_b$ $\in$ $\mathbb{C}^{{M \times 1}}$ and $\mathbf{v}_e$ $\in$ $\mathbb{C}^{{M \times 1}}$ respectively represent the transmit beamforming vectors for the CM and AN. The BS transmit power is 
	\begin{equation}
		P_{BS}=\mathbf{v}_b^H\mathbf{v}_b+\mathbf{v}_e^H\mathbf{v}_e.
	\end{equation} 
	
	The reflected signal from IRS is modeled as follows
	\begin{equation}
		\begin{aligned}
			\mathbf{y}_{i}^t&=\mathbf{\Theta}\mathbf{H}_{si}\mathbf{x}+\mathbf{\Theta}\mathbf{n_I}\\
			&=\mathbf{\Theta}\mathbf{H}_{si}(\mathbf{v}_bx_b+\mathbf{v}_ex_e)+\mathbf{\Theta}\mathbf{n}_I,
		\end{aligned}
	\end{equation}
	where  $\mathbf{\Theta}=\operatorname{diag}\left(\alpha_1e^{j\theta_1},\cdots,\alpha_Ne^{j\theta_N}\right)$ is IRS reflecting coefficient matrix. $\alpha_n$ and $\theta_n\in(0,2\pi)$ represent amplifying coefficient and phase shift of $n$-th element of active IRS, respectively. $\mathbf{n}_I$ $\in$ $\mathbb{C}^{{N \times 1}}$ with $\mathbf{n_I} \sim \mathcal{CN}\left( {0,{\mathbf{{\sigma}}^2\mathbf{I}_{{N}}}}\right)$ denotes the additive white Gaussian noise (AWGN) introduced by active IRS. 
	The reflected power at active IRS is expressed as
	\begin{equation}
		P_{i}=\|\mathbf{\Theta}\mathbf{H}_{si}(\mathbf{v}_b+\mathbf{v}_e)\|_2^2+\sigma^2\|\mathbf{\Theta}\|_F^2.		
	\end{equation}
	The received signal at Bob and Eve are
	\begin{subequations}
		\begin{align}
			&y_b=(\mathbf{h}_b^H+\mathbf{g}_b^H\mathbf{\Theta}\mathbf{H}_{si})(\mathbf{v}_bx_b+\mathbf{v}_ex_e)+\mathbf{g}_b^H\mathbf{\Theta}\mathbf{n}_I+n_b,\\
			&y_e=(\mathbf{h}_e^H+\mathbf{g}_e^H\mathbf{\Theta}\mathbf{H}_{si})(\mathbf{v}_bx_b+\mathbf{v}_ex_e)+\mathbf{g}_e^H\mathbf{\Theta}\mathbf{n}_I+n_e,
		\end{align}
	\end{subequations}
	where $n_b \in \mathbb{C}^{{1 \times 1}}$ and $n_e \in \mathbb{C}^{{1 \times 1}}$ are the AWGN, respectively. $n_b$ and $n_e$ follow Gaussian distribution of $n_b$ $\sim \mathcal{CN}\left( {0,{\sigma_b^2}}\right)$ and $n_e$ $\sim \mathcal{CN}\left( {0,{\sigma_e^2}}\right)$.
	
	The SNR at Bob and Eve are
	\begin{subequations}
		\begin{align}
			&\text{SNR}_b=\frac{|\mathbf{t}_b^H\mathbf{v}_b|^2}{|\mathbf{t}_b^H\mathbf{v}_e|^2+\sigma^2\|\mathbf{g}_b^H\mathbf{\Theta}\|_2^2+\sigma_b^2}, \\
			&\text{SNR}_e=\frac{|\mathbf{t}_e^H\mathbf{v}_b|^2}{|\mathbf{t}_e^H\mathbf{v}_e|^2+\sigma^2\|\mathbf{g}_e^H\mathbf{\Theta}\|_2^2+\sigma_e^2},
		\end{align} 
	\end{subequations}
	where  $\mathbf{t}_b^H=\mathbf{h}_b^H+\mathbf{g}_b^H\mathbf{\Theta}\mathbf{H}_{si}$ and $\mathbf{t}_e^H=\mathbf{h}_e^H+\mathbf{g}_e^H\mathbf{\Theta}\mathbf{H}_{si}$.
	
	The SR in this case is
	\begin{equation}
		\rm SR=\rm log_2(1+\text{SNR}_b)-log_2(1+\text{SNR}_b).
	\end{equation}
Given this, the maximization of SR can be expressed as 
	\begin{subequations}
		\begin{align}
			&\max_{\mathbf{v}_b,\mathbf{v}_e,\mathbf{\Theta}}~~{\rm SR}\label{SR}\\
			&~~~\text{s.t.}~~~~~\mathbf{v}_b^H\mathbf{v}_b+\mathbf{v}_e^H\mathbf{v}_e+||\mathbf{\Theta}\mathbf{H}_{si}\mathbf{v}_b||_2^2+\|\mathbf{\Theta}\mathbf{H}_{si}\mathbf{v}_e\|_2^2\nonumber\\
			&~~~~~~~~~~~+\sigma^2\|\mathbf{\Theta}\|_F^2 \le P_{max}.
		\end{align}
	\end{subequations}

	\section{The Proposed LDT-CFFP method}
	
	In order to better solve the optimization problem (\ref{SR}), LDT-CFFP method is proposed. LDT method is firstly applied to transform it into a more manageable form. Then CFFP algorithm is utilized to achieve the CF expressions of CM beamforming vector $\mathbf{v}_b$, AN beamforming vector $\mathbf{v}_e$ and active IRS phase shift matrix $\mathbf{\Theta}$. After that, iterating these variables alternately to obtain maximum SR.
	
	\subsection{Problem Reformulation}	
	
	Let us define the virtual SNR at eve is
	\begin{equation}
		\text{SNR}_E=\frac{|\mathbf{t}_e^H\mathbf{v}_e|^2}{|\mathbf{t}_e^H\mathbf{v}_b|^2+\sigma^2\|\mathbf{g}_e^H\mathbf{\Theta}\|_2^2+\sigma_e^2},
	\end{equation}
	so that the virtual SR can be written as
	\begin{equation}
		\rm VR=\rm log_2(1+\text{SNR}_b)+log_2(1+\text{SNR}_E).
	\end{equation}
	Accordingly, the optimization problem can be represented as
	\begin{subequations}
		\begin{align}
			&\max_{\mathbf{v}_b,\mathbf{v}_e,\mathbf{\Theta}}~~{\rm VR}\label{P0}\\
			&~~~\text{s.t.}~~~~~\mathbf{v}_b^H\mathbf{v}_b+\mathbf{v}_e^H\mathbf{v}_e+||\mathbf{\Theta}\mathbf{H}_{si}\mathbf{v}_b||_2^2+\|\mathbf{\Theta}\mathbf{H}_{si}\mathbf{v}_e\|_2^2\nonumber\\
			&~~~~~~~~~~+\sigma^2\|\mathbf{\Theta}\|_F^2 \le P_{max}.
		\end{align}
	\end{subequations}

	Here, some auxiliary variables  $\lambda_b \in \mathbb{R}_+$, $\lambda_e \in$ $\mathbb{R}_+$, $\mu_b \in \mathbb{C}$ and $\mu_e \in \mathbb{C}$ are introduced to the above problem, while LDT method is adopted to transform it into the following form 
	\begin{subequations}
		\begin{align}\label{P1}
			&\max_{\lambda_b,\lambda_e,\mu_b,\mu_e,\mathbf{v}_b,\mathbf{v}_e,\mathbf{\Theta}}\rm VR'\\
			&~~~~~~~~~\text{s.t.}~~~~~\mathbf{v}_b^H\mathbf{v}_b+\mathbf{v}_e^H\mathbf{v}_e+\|\mathbf{\Theta}\mathbf{H}_{si}\mathbf{v}_b\|_2^2+\|\mathbf{\Theta}\mathbf{H}_{si}\mathbf{v}_e\|_2^2\nonumber\\
			&~~~~~~~~~~~~~~~~+\sigma^2||\mathbf{\Theta}||_F^2 \le P_{max},
		\end{align}
	\end{subequations}
	where 
	\begin{align}
		&\text{VR}'=\rm ln(1+\lambda_b)+\rm ln(1+\lambda_e)-\lambda_b-\lambda_e\\
		&~~~~ -|\mu_b|^2(|\mathbf{t}_b^H\mathbf{v}_b|^2+|\mathbf{t}_b^H\mathbf{v}_e|^2+\sigma^2||\mathbf{g}_b^H\mathbf{\Theta}||_2^2+\sigma_b^2)\nonumber\\
		&~~~~ -|\mu_e|^2(|\mathbf{t}_e^H\mathbf{v}_e|^2+|\mathbf{t}_e^H\mathbf{v}_b|^2+\sigma^2||\mathbf{g}_e^H\mathbf{\Theta}||_2^2+\sigma_e^2)\nonumber\\
		&~~~~ +2\sqrt{(1+\lambda_b)} \Re\left\{\mu_b^*\mathbf{t}_b^H\mathbf{v}_b\right\}+2\sqrt{(1+\lambda_e)} \Re\left\{\mu_e^*\mathbf{t}_e^H\mathbf{v}_e\right\}. \nonumber
	\end{align}

	\subsection{Optimization of $\lambda_b $ and $\lambda_e$}

	Fixing  $\lambda_e$, $\mu_b$, $\mu_e$, $\mathbf{v}_b$, $\mathbf{v}_e$ and $\mathbf{\Theta}$, , $\lambda_b$ can be derived via $\frac{\partial \rm VR'}{\partial \rm \lambda_b}$ =0 as
	\begin{equation}
		\lambda_b^{opt}=\frac{t_b^2+t_b\sqrt{t_b^2+4}}{2},\label{lamdab}
	\end{equation}
	where $t_b$ =$\Re\left\{\mu_b^*\mathbf{t}_b^H\mathbf{v}_b\right\}$.
	
	Fixing  $\lambda_b$, $\mu_b$, $\mu_e$, $\mathbf{v}_b$, $\mathbf{v}_e$ and $\mathbf{\Theta}$, let $\frac{\partial \rm VR'}{\partial \rm \lambda_e}$ =0, thus	$\lambda_e$ can be calculated as
	\begin{equation}
		\lambda_e^{opt}=\frac{t_e^2+t_e\sqrt{t_e^2+4}}{2},\label{lamdae}
	\end{equation}where $t_e$ =$\Re\left\{\mu_e^*\mathbf{t}_e^H\mathbf{v}_e\right\}$.
	\subsection{Optimization of $\mu_b $ and $\mu_e$}
	When variable $\lambda_b$, $\lambda_e$, $\mu_e$, $\mathbf{v}_b$, $\mathbf{v}_e$ and $\mathbf{\Theta}$,  are given, we concentrate on optimizing $\mu_b$. Let $\frac{\partial \rm VR'}{\partial \rm \mu_b}$ =0, we have
	\begin{equation}
		\mu_b^{opt}=\frac{\sqrt{1+\lambda_b}\mathbf{t}_b^H\mathbf{v}_b}{|\mathbf{t}_b^H\mathbf{v}_b|^2+|\mathbf{t}_b^H\mathbf{v}_e|^2+\sigma^2\|\mathbf{g}_b^H\mathbf{\Theta}\|_2^2+\sigma_b^2}.\label{mub}
	\end{equation}
	
	Given variable $\lambda_b$, $\lambda_e$, $\mu_b$, $\mathbf{v}_b$, $\mathbf{v}_e$ and $\mathbf{\Theta}$,  we pay attention to optimize $\mu_e$. Let $\frac{\partial \rm VR'}{\partial \rm \mu_e}$ =0, we obtain 
	\begin{equation}
		\mu_e^{opt}=\frac{\sqrt{1+\lambda_e}\mathbf{t}_e^H\mathbf{v}_e}{|\mathbf{t}_e^H\mathbf{v}_e|^2+|\mathbf{t}_e^H\mathbf{v}_b|^2+\sigma^2\|\mathbf{g}_e^H\mathbf{\Theta}\|_2^2+\sigma_e^2}.\label{mue}
	\end{equation}
	\subsection{Optimization of beamforming vector ${\mathbf{v}}_b$ and ${\mathbf{v}}_e$}
	Given $\lambda_b$, $\lambda_e$, $\mu_b$, $\mu_e$, $\mathbf{v}_e$ and $\mathbf{\Theta}$,   our attention turns to optimizing $\mathbf{v}_b$. The problem (\ref{P0}) is transformed into
	\begin{subequations}
		\begin{align}
			&\max_{\mathbf{v}_b}~~~~\Re\left\{2\mathbf{a}_b^H\mathbf{v}_b\right\}-\mathbf{v}_b^H\mathbf{A}\mathbf{v}_b\label{vb}\\
			&~~\text{s.t.}~~~~~~\mathbf{v}_b^H\mathbf{F}\mathbf{v}_b \le P_b,
		\end{align}
	\end{subequations}
	where 
	\begin{subequations}
		\begin{align}
			&\mathbf{a}_b^H=\sqrt{(1+\lambda_b)}\mu_b^*\mathbf{t}_b^H,\\
			&\mathbf{A}=|\mu_b|^2\mathbf{t}_b\mathbf{t}_b^H+|\mu_e|^2\mathbf{t}_e\mathbf{t}_e^H,\\
			&\mathbf{F}=\mathbf{I}_K\otimes(\mathbf{I}+\mathbf{H}_{si}^H\mathbf{\Theta}^H\mathbf{\Theta}\mathbf{H}_{si}),\\
			&P_b=P_{max}-\mathbf{v}_e^H\mathbf{v}_e-\|\mathbf{\Theta}\mathbf{H}_{si}\mathbf{v}_e\|_2^2-\sigma^2\|\mathbf{\Theta}\|_F^2.
		\end{align}
	\end{subequations}
	
	
	Fixing $\lambda_b$, $\lambda_e$, $\mu_b$, $\mu_e$, $\mathbf{v}_b$ and $\mathbf{\Theta}$,  we focus on the optimization of $\mathbf{v}_e$. The problem (\ref{P0}) is transformed into
	\begin{subequations}
		\begin{align}
			&\max_{\mathbf{v}_e}~~~~\Re\left\{2\mathbf{a}_e^H\mathbf{v}_e\right\}-\mathbf{v}_e^H\mathbf{A}\mathbf{v}_e\label{ve}\\
			&~~\text{s.t.}~~~~~~\mathbf{v}_e^H\mathbf{F}\mathbf{v}_e \le P_e,
		\end{align}
	\end{subequations}		
	where 
	\begin{subequations}
		\begin{align}
			&\mathbf{a}_e^H=\sqrt{(1+\lambda_e)}\mu_e^*\mathbf{t}_e^H,\\
			&P_e=P_{max}-\mathbf{v}_b^H\mathbf{v}_b-\|\mathbf{\Theta}\mathbf{G}\mathbf{v}_b\|_2^2-\sigma^2\|\mathbf{\Theta}\|_F^2.
		\end{align}
	\end{subequations}
	
	Clearly, problem (\ref{vb}) and (\ref{ve}) meet the convex property, thus they can be directly solved by optimization tools like CVX.
	\subsection{Optimization of IRS reflecting coefficient matrix $\mathbf{\Theta}$} 
	With variable $\lambda_b$, $\lambda_e$, $\mu_b$, $\mu_e$, $\mathbf{v}_b$ and $\mathbf{v}_e$,  fixed, we optimize $\mathbf{\Theta}$. Let $\boldsymbol{\theta}=(a_1e^{j\theta_1},\cdots,a_Ne^{j\theta_N})^H$, we have the following transformation: $\mathbf{g}^H\mathbf{\Theta}=\boldsymbol{\theta}^H\operatorname{diag}(\mathbf{g}^H)$,  $\|\mathbf{\Theta}\|_F^2=\boldsymbol{\theta}^H\boldsymbol{\theta}$ and $\mathbf{\Theta}\mathbf{H}_{si}\mathbf{v}=\operatorname{diag}(\mathbf{H}_{si}\mathbf{v})\boldsymbol{\theta}^*$, thus the optimization problem (\ref{P1}) can be rewritten as
	\begin{subequations}
		\begin{align}
			&\max_{\boldsymbol{\theta}}~~~~\Re\left\{2\boldsymbol{\theta}^H\boldsymbol{\chi}\right\}-\boldsymbol{\theta}^H\mathbf{\Upsilon}\boldsymbol{\theta}\label{theta}\\
			&~~\text{s.t.}~~~~~~\boldsymbol{\theta}^H\mathbf{\Omega}\boldsymbol{\theta} \le P_{be},
		\end{align}
	\end{subequations}

	where
	\begin{subequations}
		\begin{align}
			&P_{be}=P_{max}-\mathbf{v}_b^H\mathbf{v}_b-\mathbf{v}_e^H\mathbf{v}_e,\\
			&\mathbf{\Upsilon}=	
			|\mu_b|^2\operatorname{diag}\left(\mathbf{g}_b^H\right)\mathbf{H}_{si}\mathbf{v}_b\mathbf{v}_b^H\mathbf{H}_{si}^H(\operatorname{diag}\left(\mathbf{g}_b^H\right))^H,\\
			&\mathbf{\Omega}=|\mu_b|^2\operatorname{diag}\left(\mathbf{g}_b^H\right)\mathbf{H}_{si}\mathbf{v}_e\mathbf{v}_e^H\mathbf{H}_{si}^H(\operatorname{diag}\left(\mathbf{g}_b^H\right))^H\\
			&~~~~+|\mu_e|^2\operatorname{diag}\left(\mathbf{g}_e^H\right)\mathbf{H}_{si}\mathbf{v}_b\mathbf{v}_b^H\mathbf{H}_{si}^H(\operatorname{diag}\left(\mathbf{g}_e^H\right))^H\nonumber\\
			&~~~~+|\mu_e|^2\operatorname{diag}\left(\mathbf{g}_e^H\right)\mathbf{H}_{si}\mathbf{v}_e\mathbf{v}_e^H\mathbf{H}_{si}^H(\operatorname{diag}\left(\mathbf{g}_e^H\right))^H\nonumber\\
			&~~~~+\sigma^2|\mu_b|^2\operatorname{diag}\left(\mathbf{g}_b^H\right)(\operatorname{diag}\left(\mathbf{g}_b^H\right))^H\nonumber\\
			&~~~~+\sigma^2|\mu_e|^2\operatorname{diag}\left(\mathbf{g}_e^H\right)(\operatorname{diag}\left(\mathbf{g}_e^H\right))^H\nonumber\\
			&~~~~+\operatorname{diag}\left(\mathbf{H}_{si}\mathbf{v}_b\right)(\operatorname{diag}\left(\mathbf{H}_{si}\mathbf{v}_b\right))^H\nonumber\\
			&~~~~+\operatorname{diag}\left(\mathbf{H}_{si}\mathbf{v}_e\right)(\operatorname{diag}\left(\mathbf{H}_{si}\mathbf{v}_e\right))^H+\sigma^2\mathbf{I},\nonumber\\		
			&\boldsymbol{\chi}=\sqrt{(1+\lambda_b)}\operatorname{diag}\left(\mu_b^*\mathbf{g}_b^H\right)\mathbf{H}_{si}\mathbf{v}_b\\					&~~~~+\sqrt{(1+\lambda_e)}\operatorname{diag}\left(\mu_e^*\mathbf{g}_e^H\right)\mathbf{H}_{si}\mathbf{v}_e\nonumber\\
			&~~~~-|\mu_b|^2\operatorname{diag}\left(\mathbf{g}_b^H\right)\mathbf{H}_{si}\mathbf{v}_b\mathbf{v}_b^H\mathbf{h}_b\nonumber\\
			&~~~~-|\mu_b|^2\operatorname{diag}\left(\mathbf{g}_b^H\right)\mathbf{H}_{si}\mathbf{v}_e\mathbf{v}_e^H\mathbf{h}_b\nonumber\\
			&~~~~-|\mu_e|^2\operatorname{diag}\left(\mathbf{g}_e^H\right)\mathbf{H}_{si}\mathbf{v}_e\mathbf{v}_e^H\mathbf{h}_e\nonumber\\
			&~~~~-|\mu_e|^2\operatorname{diag}\left(\mathbf{g}_e^H\right)\mathbf{H}_{si}\mathbf{v}_b\mathbf{v}_b^H\mathbf{h}_e.\nonumber
		\end{align}
	\end{subequations}
	
	Problem (\ref{theta}) is clearly convex, and its solution $\boldsymbol{\theta}$ can be directly acquired through CVX. Afterwards, IRS reflecting coefficient matrix $\mathbf{\Theta}$ is achieved.
	
	The computational complexity of LDT-CFFP algorithm is $\mathcal{O}\{ I[\sqrt{KM}(2K^5M^5+K^4M^4)+\sqrt{N}(2N^4+N^3)+K^2M+KN+KM]\}$, where $I$ is the number of internal loops in LDT-CFFP algorithm and $K$ is the number of uesrs.

	\begin{algorithm}
		\caption{ Proposed LDT-CFFP Algorithm}
		\begin{algorithmic}[1]
			\STATE Initialize $\mathbf{v}$ and $\boldsymbol{\theta}$.
			\STATE Set i=1.
			\REPEAT
			\STATE Update $\lambda_b(i)$ and $\lambda_e(i)$ by (\ref{lamdab}) and (\ref{lamdae}).
			\STATE Update $\mu_b(i)$ and $\mu_e(i)$ by (\ref{mub}) and (\ref{mue}).
			\STATE Update $\mathbf{v}_b(i)$ and $\mathbf{v}_e(i)$ by (\ref{vb}) and (\ref{ve}).
			\STATE Update $\boldsymbol{\theta}(i)$ by (\ref{theta}).
			\STATE Calculate $\rm VR'(i)$.
			\STATE Set i=i+1.
			\UNTIL $|\rm VR'(i+1) - \rm VR'(i)| \le 10^{-4}$.
			\STATE  $\rm VR'(i)$ is the optimal solution of (\ref{P1}). 
			Afterwards, we will incorporate $\mathbf{v}_b(i)$, $\mathbf{v}_e(i)$ and $\boldsymbol{\theta}(i)$ into (\ref{SR}) to obtain the solution of SR.
		\end{algorithmic}
	\end{algorithm}
	\section{The Proposed NSP-MRR-PA method }
	
	To reduce the high computational complexity of LDT-CFFP method ($\mathcal{O}\{ M^{5.5}+N^{4.5} \}$), a block-structured IRS-assisted DM network is proposed. Subsequently, a low-complexity NSP-MRR-PA method is proposed. By NSP method, IRS blocks can respectively serve for Bob and Eve. Besides, we adopt three PA schemes, i.e., ES \cite{NA9780538733519}, PSO \cite{HL10361533}, SA \cite{ZJH10284767}, to boost SR. 
	
	\subsection{Blocked Active IRS-Assisted DM Wireless System}
	
	\begin{figure}
		\centering
		\includegraphics[width=0.5\textwidth]{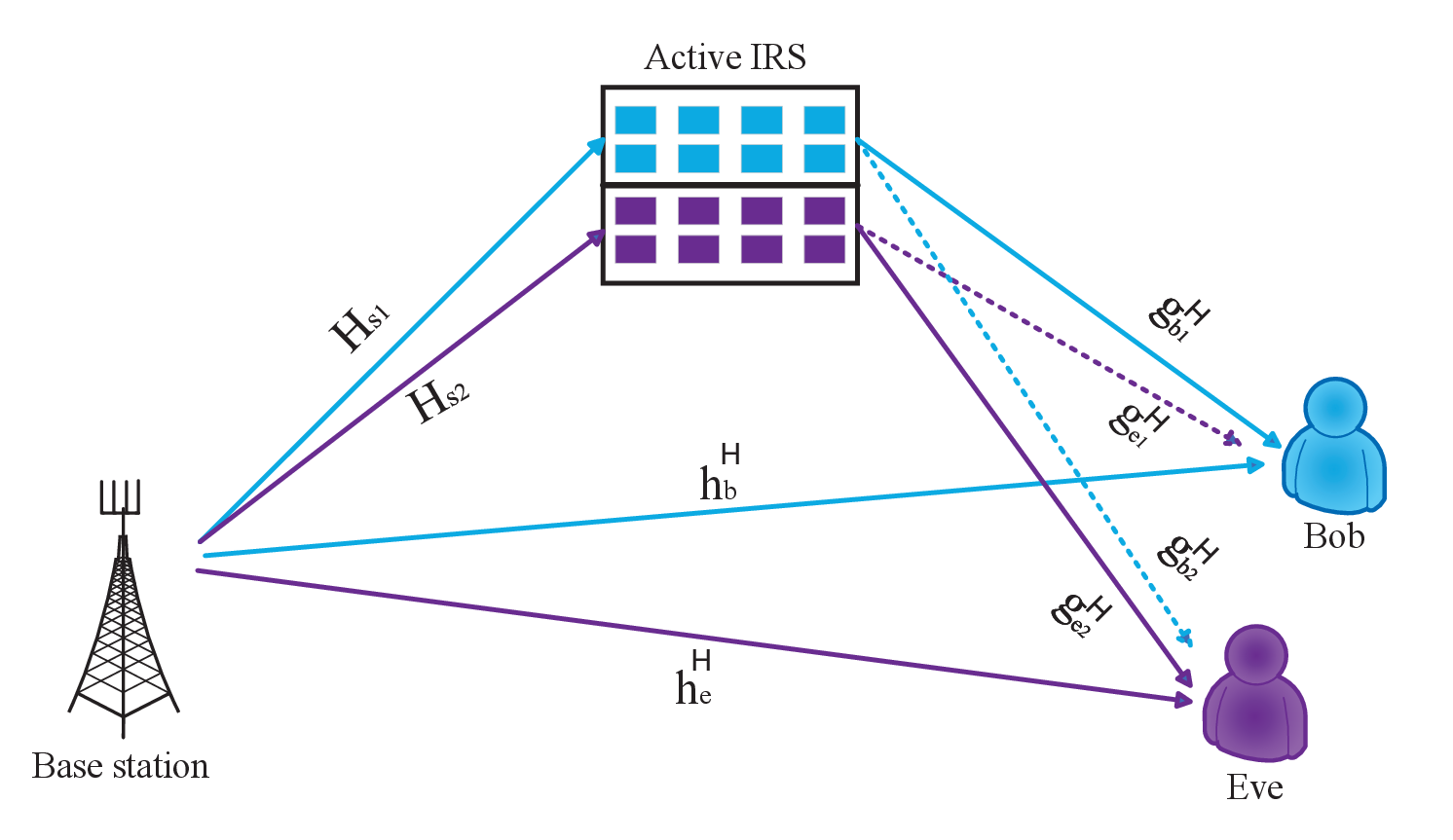}
		\caption{ Blocked active IRS-assisted DM secure communication system}
		\label{block}
	\end{figure}
	As shown in Fig. \ref{block}, we divide the active IRS into two parts, and each part is with $N_k$ reflective elements, where $k=\{1,2\}$. Here, we consider PA strategies between BS and IRS, between CM and AN. 
	The transmit signal at BS is
	\begin{equation}
		\mathbf{x}=\sqrt{\eta\beta P_s}\widetilde{\mathbf{v}}_bx_b+\sqrt{\eta(1-\beta)P_s}\widetilde{\mathbf{v}}_ex_e,
	\end{equation}
	where $P_s$ is the total power composed of BS transmit power and the reflected power via IRS. $\widetilde{\mathbf{v}}_b\in\mathbb{C}^{M\times 1}$ and $\widetilde{\mathbf{v}}_e\in\mathbb{C}^{M\times 1}$ are the beamforming vectors for CM and AN, where $\|\widetilde{\mathbf{v}}_b\|^2=\|\widetilde{\mathbf{v}}_e\|^2=1$,  respectively. Additionally, $\eta\in(0,1)$ and $\beta\in(0,1)$ are the PA parameters of BS and IRS, CM and AN separately.
	
	The reflected signal from $k$-th IRS block is
	\begin{align}
		\mathbf{y}_{ik}^t&=\mathbf{\Theta}_k(\mathbf{H}_{sk}\mathbf{x}+\mathbf{n}_{k})\\
		&=\mathbf{\Theta}_k\mathbf{H}_{sk}(\sqrt{\eta\beta P_s}\widetilde{\mathbf{v}}_bx_b+\sqrt{\eta(1-\beta)P_s}\widetilde{\mathbf{v}}_ex_e)+\mathbf{\Theta}_k\mathbf{n}_k, \nonumber
	\end{align}
	where $\mathbf{H}_{sk}\in\mathbb{C}^{N_k\times M}$ represents the channel from BS to $k$-th block, $\mathbf{\Theta}_k=\text{diag}(\alpha_{k1} e^{j\theta_{k1}}, \cdots, \alpha_{kN_k} e^{j\theta_{kN_k}}) \in\mathbb{C}^{N_k\times N_k}$ denotes the reflecting coefficient matrix of $k$-th block, where $\alpha_{kn}$ and $\theta_{kn}\in(0,2\pi)$ stand for amplifying coefficients of the $n$-th element in $k$-th block. $\mathbf{n}_k \in\mathbb{C}^{N_k\times 1}$ is the AWGN generated by active IRS, it subject to the distribution of $\mathbf{n}_k\sim\mathcal{C}\mathcal{N}(0,~\sigma_k^2\mathbf{I}_{N_k})$.

	The signals received at Bob and Eve can be respectively represented as
	\begin{subequations}
		\begin{align}
			&y_b=(\mathbf{h}_b^H+\sum_{k=1}^{2} \mathbf{g}_{bk}^H\mathbf{\Theta}_k\mathbf{H}_{sk})\mathbf{x}+\sum_{k=1}^{2} \mathbf{g}_{bk}^H\mathbf{\Theta}_k\mathbf{n}_{k}+n_b, \\
			&y_e=(\mathbf{h}_b^H+\sum_{k=1}^{2} \mathbf{g}_{ek}^H\mathbf{\Theta}_k\mathbf{H}_{sk})\mathbf{x}+\sum_{k=1}^{2} \mathbf{g}_{ek}^H\mathbf{\Theta}_k\mathbf{n}_{k}+n_e, 
		\end{align}
	\end{subequations}
	where $\mathbf{g}_{bk}^H\in\mathbb{C}^{1\times N_k}$ and $\mathbf{g}_{ek}^H\in\mathbb{C}^{1\times N_k}$ denote the channels from $k$-th IRS block to Bob and to Eve, respectively. $\mathbf{h}_b^H\in\mathbb{C}^{1\times M}$ and $\mathbf{h}_e^H\in\mathbb{C}^{1\times M}$ represent the channels from from BS to Bob and to Eve, respectively. Afterwards, let
	$\boldsymbol{\theta}_k=\rho_k\widetilde{\boldsymbol{\theta}}_k=[\alpha_{k1} e^{j\theta_{k1}}, \cdots, \alpha_{kN_k} e^{j\theta_{kN_k}}] ^H$,  $\rho_k=\|\boldsymbol{\theta}_k\|_2$,   $\|\widetilde{\boldsymbol{\theta}}_k\|_2=1$, $\mathbf{G}_{bk}=\text{diag}(\mathbf{g}_{bk}^H)$ and   $\mathbf{G}_{ek}=\text{diag}(\mathbf{g}_{ek}^H)$. Consequently, $y_b$ and $y_e$ can be rewritten as
	\begin{subequations}
		\begin{align}
			y_b&=\sqrt{\eta\beta P_s}(\mathbf{h}_b^H+\sum_{k=1}^{2}\rho_k\widetilde{\boldsymbol{\theta}}_k^H\mathbf{G}_{bk}\mathbf{H}_{sk})\widetilde{\mathbf{v}}_bx_b \label{y_b}\\
			&+\sqrt{\eta(1-\beta)P_s}(\mathbf{h}_b^H+\sum_{k=1}^{2}\rho_k\widetilde{\boldsymbol{\theta}}_k^H\mathbf{G}_{bk}\mathbf{H}_{sk})\widetilde{\mathbf{v}}_ex_e\nonumber\\
			&+\sum_{k=1}^{2}\rho_k\widetilde{\boldsymbol{\theta}}_k^H\mathbf{G}_{bk}\mathbf{n}_{k}+n_b,  \nonumber\\
			y_e	&=\sqrt{\eta\beta P_s}(\mathbf{h}_e^H+\sum_{k=1}^{2}\rho_k\widetilde{\boldsymbol{\theta}}_k^H\mathbf{G}_{ek}\mathbf{H}_{sk})\widetilde{\mathbf{v}}_bx_b\label{y_e}\\
			&+\sqrt{\eta(1-\beta)P_s}(\mathbf{h}_e^H+\sum_{k=1}^{2}\rho_k\widetilde{\boldsymbol{\theta}}_k^H\mathbf{G}_{ek}\mathbf{H}_{sk})\widetilde{\mathbf{v}}_ex_e\nonumber\\
			&+\sum_{k=1}^{2}\rho_k\widetilde{\boldsymbol{\theta}}_k^H\mathbf{G}_{ek}\mathbf{n}_{k}+n_e, \nonumber
		\end{align}
	\end{subequations}
	
	The SNRs at Bob and Eve are denoted as
	\begin{subequations}
		\begin{align}
			\hat{\gamma}_b=\frac{M_1}{M_2+\sum_{k=1}^{2}\sigma_k^2\rho_k^2\|\widetilde{\boldsymbol{\theta}}_k^H\mathbf{G}_{bk}\|^2+\sigma_b^2},\label{gammab} \\
			\hat{\gamma}_e=\frac{\hat{M}_2}{\hat{M}_1+\sum_{k=1}^{2}\sigma_k^2\rho_k^2\|\widetilde{\boldsymbol{\theta}}_k^H\mathbf{G}_{ek}\|^2+\sigma_e^2},\label{gammae}
		\end{align}
	\end{subequations}
	
	where
	\begin{subequations}
		\begin{align}
			&M_1=\eta\beta P_s\|(\mathbf{h}_b^H+\sum_{k=1}^{2}\rho_k\widetilde{\boldsymbol{\theta}}_k^H\mathbf{G}_{bk}\mathbf{H}_{sk})\widetilde{\mathbf{v}}_b\|^2,\\
			&M_2=\eta(1-\beta) P_s\|(\mathbf{h}_b^H+\sum_{k=1}^{2}\rho_k\widetilde{\boldsymbol{\theta}}_k^H\mathbf{G}_{bk}\mathbf{H}_{sk})\widetilde{\mathbf{v}}_e\|^2,\\
			&\hat{M}_1=\eta(1-\beta) P_s\|(\mathbf{h}_e^H+\sum_{k=1}^{2}\rho_k\widetilde{\boldsymbol{\theta}}_k^H\mathbf{G}_{ek}\mathbf{H}_{sk})\widetilde{\mathbf{v}}_e\|^2,\\
			&\hat{M}_2=\eta\beta P_s\|(\mathbf{h}_e^H+\sum_{k=1}^{2}\rho_k\widetilde{\boldsymbol{\theta}}_k^H\mathbf{G}_{ek}\mathbf{H}_{sk})\widetilde{\mathbf{v}}_b\|^2.
		\end{align}
	\end{subequations} 
	
	Similarly, the SR of blocked IRS-assisted DM secure wireless system is given by
	\begin{equation}
		\hat{\text{SR}}=\log_2(1+\hat{\gamma}_b)-\log_2(1+\hat{\gamma}_e).
	\end{equation}
	As a result, the corresponding optimization problem is modeled as follows
	\begin{subequations}
		\begin{align}
			&\max_{\eta,\beta,\rho_k,\widetilde{\boldsymbol{\theta}}_k,\widetilde{\mathbf{v}}_b,\widetilde{\mathbf{v}}_e}~~{\rm SR}_1\label{SR11}\\
			&~~~~~~~\text{s.t.}~~~~~~~~
			0\le \eta \le 1,~0\le\beta\le 1,\\
			&~~~~~~~~~~~~~~~~~~\rho_k=\|\boldsymbol{\theta}_k\|_2,~\|\widetilde{\boldsymbol{\theta}}_k\|^2=1,\\
			&~~~~~~~~~~~~~~~~~~\|\widetilde{\mathbf{v}}_b\|_2^2=\|\widetilde{\mathbf{v}}_e\|_2^2=1.
		\end{align}
	\end{subequations}

	\subsection{Optimizating beamforming vector $\widetilde{\mathbf{v}}_b$, $\widetilde{\mathbf{v}}_e$}
	To eliminate the interference of AN on Bob, NSP method is operated to project AN onto the null space of $\mathbf{h}_b^H$ and $\mathbf{H}_{s1}$. When $\eta$, $\beta$, $\rho_k$ and $\widetilde{\boldsymbol{\theta}}_k$ are fixed, in accordance with \cite{LYQ10141981}, the optimization problem related to  $\widetilde{\mathbf{v}}_e$ is 
	\begin{subequations}
		\begin{align}	&\max_{\widetilde{\mathbf{v}}_e}~~\widetilde{\mathbf{v}}_e^H(\mathbf{h}_b^H+\rho_2\widetilde{\boldsymbol{\theta}}_2^H\mathbf{G}_{b2}\mathbf{H}_{s2})^H(\mathbf{h}_b^H+\rho_2\widetilde{\boldsymbol{\theta}}_2^H\mathbf{G}_{b2}\mathbf{H}_{s2})\widetilde{\mathbf{v}}_e \label{zero_ve}\\
			&~~\text{s.t.}~~~~
			(\mathbf{h}_b~ \mathbf{H}_{s1}^H)^H\widetilde{\mathbf{v}}_e=\mathbf{0},~\widetilde{\mathbf{v}}_e^H\widetilde{\mathbf{v}}_e=1.
		\end{align}	
	\end{subequations}
	where 
	$\widetilde{\mathbf{v}}_e=\mathbf{T}_1\cdot\mathbf{u}_e$. Let 
	$\mathbf{Q}_1=(\mathbf{h}_b~\mathbf{H}^H_{s1})^H$, we have  $\mathbf{T}_1=[\mathbf{I}_M-\mathbf{Q}_1^H(\mathbf{Q}_1\mathbf{Q}_1^H)^{\dagger}\mathbf{Q}_1]$. Thus, problem (\ref{zero_ve}) is updated to
	\begin{subequations}
		\begin{align}	&\max_{\mathbf{u}_e}~~~\mathbf{u}_e^H\mathbf{T}_1^H\widetilde{\mathbf{h}}_1^H\widetilde{\mathbf{h}}_1\mathbf{T}_1\mathbf{u}_e \label{zero_ue}\\
			&~\text{s.t.}~~~~~
			\mathbf{u}_e^H\mathbf{u}_e=1,
		\end{align}	
	\end{subequations}
	where    $\widetilde{\mathbf{h}}_1=\mathbf{h}_b^H+\rho_2\widetilde{\boldsymbol{\theta}}_2^H\mathbf{G}_{b2}\mathbf{H}_{s2}$. 
	Through (\ref{zero_ue}), it can be obtained that $\text{Rank}(\mathbf{T}_1^H\widetilde{\mathbf{h}}_1^H\widetilde{\mathbf{h}}_1\mathbf{T}_1)=1$. Therefore, $\widetilde{\mathbf{v}}_e$ is presented as follows
	\begin{equation}
		\widetilde{\mathbf{v}}_e=\frac{\mathbf{T}_1\widetilde{\mathbf{h}}_1^H}{\|\mathbf{T}_1\widetilde{\mathbf{h}}_1^H\|}.\label{v_e}
	\end{equation}
	Subsequently, the way to solve $\widetilde{\mathbf{v}}_b$ is similar to the method for solving $\widetilde{\mathbf{v}}_e$. For solving $\widetilde{\mathbf{v}}_e$, CM is projected onto the null space of $\mathbf{h}_e^H$ and $\mathbf{H}_{s2}$, while the optimization problem is written as follows
	\begin{subequations}
		\begin{align}	&\max_{\widetilde{\mathbf{v}}_b}~~\widetilde{\mathbf{v}}_b^H(\mathbf{h}_e^H+\rho_2\widetilde{\boldsymbol{\theta}}_2^H\mathbf{G}_{e2}\mathbf{H}_{s2})^H(\mathbf{h}_e^H+\rho_2\widetilde{\boldsymbol{\theta}}_2^H\mathbf{G}_{e2}\mathbf{H}_{s2})\widetilde{\mathbf{v}}_b \label{zero_vb}\\
			&~~\text{s.t.}~~~~
			(\mathbf{h}_e~ \mathbf{H}_{s2}^H)\widetilde{\mathbf{v}}_b=0,~\widetilde{\mathbf{v}}_b^H\widetilde{\mathbf{v}}_b=1.
		\end{align}	
	\end{subequations}
	We obtain $\widetilde{\mathbf{v}}_b$ as follows
	\begin{equation}
		\widetilde{\mathbf{v}}_b=\frac{\mathbf{T}_2\widetilde{\mathbf{h}}_2^H}{\|\mathbf{T}_2\widetilde{\mathbf{h}}_2^H\|},\label{v_b}
	\end{equation}
	where $\widetilde{\mathbf{h}}_2=\mathbf{h}_e^H+\rho_2\widetilde{\boldsymbol{\theta}}_2^H\mathbf{G}_{e2}\mathbf{H}_{s2}$, $\mathbf{T}_2=[\mathbf{I}_M-\mathbf{Q}_2^H(\mathbf{Q}_2\mathbf{Q}_2^H)^{\dagger}\mathbf{Q}_2]$ and $\mathbf{Q}_2=[\mathbf{h}_e~\mathbf{H}^H_{s2}]^H$.
	%
	
	\subsection{Optimizating IRS reflecting coefficient vector $\widetilde{\boldsymbol{\theta}}_k$}
	For the optimization of $ \widetilde{\boldsymbol{\theta}}_k $, we use MRR algorithm with $\eta$, $\beta$, $\rho_k$, $\widetilde{\mathbf{v}}_b$ and $\widetilde{\mathbf{v}}_e$ fixed.
	According to \cite{LJ10135161}, we can get $\widetilde{\boldsymbol{\theta}}_k$ through MRR method as below
	\begin{subequations}
		\begin{align}
			&\widetilde{\boldsymbol{\theta}}_1^H=\frac{\widetilde{\mathbf{v}}_b^H\mathbf{H}_{s1}^H\mathbf{G}_{b1}^H}{\|\mathbf{G}_{b1}\mathbf{H}_{s1}\widetilde{\mathbf{v}}_b\|_2}e^{j\phi_{\mathbf{h}_b^H\widetilde{\mathbf{v}}_b}},\label{MRR_theta_b}\\
			&\widetilde{\boldsymbol{\theta}}_2^H=\frac{\widetilde{\mathbf{v}}_e^H\mathbf{H}_{s2}^H\mathbf{G}_{e2}^H}{\|\mathbf{G}_{e2}\mathbf{H}_{s2}\widetilde{\mathbf{v}}_e\|_2}e^{j\phi_{\mathbf{h}_e^H\widetilde{\mathbf{v}}_e}},\label{MRR_theta_e}
		\end{align}
	\end{subequations}
	where $\phi_{\mathbf{h}_b^H\widetilde{\mathbf{v}}_b}$ and $\phi_{\mathbf{h}_b^H\widetilde{\mathbf{v}}_e}$ are the phases of $\mathbf{h}_b^H\widetilde{\mathbf{v}}_b$ and $\mathbf{h}_e^H\widetilde{\mathbf{v}}_e$, respectively. 
	
	\subsection{Optimizating $\rho_k$}	  
	Based on the derivations of $\widetilde{\mathbf{v}}_b$ and  $\widetilde{\mathbf{v}}_e$ via NSP method, and $\widetilde{\boldsymbol{\theta}}_k$ through MRR method, the received signal (\ref{y_b}) and (\ref{y_e}) are simplified as
	\begin{subequations}
		\begin{align}
			&\widetilde{\mathbf{y}}_{i1}^t=\sqrt{\eta\beta P_s}\rho_1\widetilde{\boldsymbol{\theta}}_1^H\text{diag}(\mathbf{H}_{s1}\widetilde{\mathbf{v}}_b)x_b+\rho_1\text{diag}(\widetilde{\boldsymbol{\theta}}_1^H)\mathbf{n}_1,\\
			&\widetilde{\mathbf{y}}_{i2}^t=\sqrt{\eta(1-\beta)P_s}\rho_2\widetilde{\boldsymbol{\theta}}_2^H\text{diag}(\mathbf{H}_{s2}\widetilde{\mathbf{v}}_e)x_e+\rho_2\text{diag}(\widetilde{\boldsymbol{\theta}}_2^H)\mathbf{n}_2.
		\end{align}
	\end{subequations}
	The total reflected average power at IRS is
	\begin{align}
		P_I=&E\{(\widetilde{\mathbf{y}}_{ib}^t)^H\widetilde{\mathbf{y}}_{ib}^t\}+E\{(\widetilde{\mathbf{y}}_{ie}^t)^H\widetilde{\mathbf{y}}_{ie}^t\},\label{PI}\\
		=&\eta\beta P_s\rho_1^2\|\widetilde{\boldsymbol{\theta}}_1^H\text{diag}(\mathbf{H}_{s1}\widetilde{\mathbf{v}}_b)\|^2+\sigma_{1}^2\rho_1^2\nonumber\\
		&+\eta(1-\beta) P_s\rho_2^2\|\widetilde{\boldsymbol{\theta}}_2^H\text{diag}(\mathbf{H}_{s2}\widetilde{\mathbf{v}}_e)\|^2+\sigma_{2}^2\rho_2^2,\nonumber\\
		=&(1-\eta)P_s.\nonumber
	\end{align}
	According to (\ref{PI}), the CF solution to $\rho_k$ can be derived as below
	\begin{subequations}
		\begin{align}
			&\rho_1=\sqrt{\frac{(1-\eta)\mu P_s}{\eta\beta P_s\|\widetilde{\boldsymbol{\theta}}_1^H\text{diag}(\mathbf{H}_{s1}\widetilde{\mathbf{v}}_b)\|^2+\sigma_{1}^2}},\label{rou1}\\
			&\rho_2=\sqrt{\frac{(1-\eta)(1-\mu) P_s}{\eta(1-\beta) P_s\|\widetilde{\boldsymbol{\theta}}_2^H\text{diag}(\mathbf{H}_{s2}\widetilde{\mathbf{v}}_e)\|^2+\sigma_{2}^2}},\label{rou2}
		\end{align}
	\end{subequations}
	where $\mu$ is the PA factor of power reflected by each IRS block.  $E\{(\widetilde{\mathbf{y}}_{ib}^t)^H\widetilde{\mathbf{y}}_{ib}^t\}=\mu(1-\eta)P_s$ and $E\{(\widetilde{\mathbf{y}}_{ie}^t)^H\widetilde{\mathbf{y}}_{ie}^t\}=(1-\mu)(1-\eta)P_s$. 
	\subsection{Optimizating PA factor $\eta$, $\beta$}
	After the use of NSP method and MRR method, 
	the received signals at Bob and Eve can be redenoted as
	\begin{subequations}
		\begin{align}
			\widetilde{y}_b
			&=\sqrt{\eta\beta P_s}(\mathbf{h}_b^H+\rho_1\widetilde{\boldsymbol{\theta}}_1^H\mathbf{G}_{b1}\mathbf{H}_{s1})\widetilde{\mathbf{v}}_bx_b\label{y'_b}\\
			&+\sqrt{\eta(1-\beta)P_s}\rho_2\widetilde{\boldsymbol{\theta}}_2^H\mathbf{G}_{b2}\mathbf{H}_{s2}\widetilde{\mathbf{v}}_ex_e\nonumber\\
			&+\rho_1\widetilde{\boldsymbol{\theta}}_1^H\mathbf{G}_{b1}\mathbf{n}_{1}+\rho_2\widetilde{\boldsymbol{\theta}}_2^H\mathbf{G}_{b2}\mathbf{n}_{2}+n_b,\nonumber\\
			\widetilde{y}_e
			&=\sqrt{\eta\beta P_s}\rho_1\widetilde{\boldsymbol{\theta}}_1^H\mathbf{G}_{e1}\mathbf{H}_{s1}\widetilde{\mathbf{v}}_bx_b\label{y'_e}\\
			&+\sqrt{\eta(1-\beta) P_s}(\mathbf{h}_e^H+\rho_2\widetilde{\boldsymbol{\theta}}_2^H\mathbf{G}_{e2}\mathbf{H}_{s2})\widetilde{\mathbf{v}}_ex_e\nonumber\\
			&+\rho_1\widetilde{\boldsymbol{\theta}}_1^H\mathbf{G}_{e1}\mathbf{n}_{1}+\rho_2\widetilde{\boldsymbol{\theta}}_2^H\mathbf{G}_{e2}\mathbf{n}_{2}+n_e.\nonumber
		\end{align}
	\end{subequations}
	Accordingly, the SINRs at Bob and Eve are respectively given by
	\begin{subequations}
		\begin{align}
			&\hat{\gamma}_b=\frac{\eta\beta P_s|(\mathbf{h}_b^H+\rho_1\widetilde{\boldsymbol{\theta}}_1^H\mathbf{G}_{b1}\mathbf{H}_{s1})\widetilde{\mathbf{v}}_b|^2}{L_1+\sigma_{1}^2\rho_1^2\|\widetilde{\boldsymbol{\theta}}_1^H\mathbf{G}_{b1}\|^2+\sigma_{2}^2\rho_2^2\|\widetilde{\boldsymbol{\theta}}_2^H\mathbf{G}_{b2}\|^2+\sigma_b^2},\label{gamma_b}\\
			&\hat{\gamma}_e=\frac{ \eta\beta P_s\rho_1^2| \widetilde{\boldsymbol{\theta}}_1^H\mathbf{G}_{e1}\mathbf{H}_{s1}\widetilde{\mathbf{v}}_b|^2}{L_2+\sigma_{1}^2\rho_1^2\|\widetilde{\boldsymbol{\theta}}_{1}^H\mathbf{G}_{e1}\|^2+\sigma_{2}^2\rho_2^2\|\widetilde{\boldsymbol{\theta}}_{2}^H\mathbf{G}_{e2}\|^2+\sigma_e^2},\label{gamma_e}
		\end{align}
	\end{subequations}
	where
	\begin{subequations}
		\begin{align}
			&L_1=\eta(1-\beta)P_s\rho_2^2|\widetilde{\boldsymbol{\theta}}_2^H\mathbf{G}_{b2}\mathbf{H}_{s2}\widetilde{\mathbf{v}}_e|^2,\\
			&L_2=\eta(1-\beta)P_s|(\mathbf{h}_e^H+\rho_2\widetilde{\boldsymbol{\theta}}_2^H\mathbf{G}_{e2}\mathbf{H}_{s2})\widetilde{\mathbf{v}}_e|^2.
		\end{align}
	\end{subequations}
	For the convenience of solving $\eta$ and $\beta$, new variables are defined as follow
	\begin{subequations}
		\begin{align}	
			&a=|\widetilde{\boldsymbol{\theta}}_1^H\mathbf{G}_{b1}\mathbf{H}_{s1}\widetilde{\mathbf{v}}_b|^2,\\
			&b=\mathfrak{R} \{\widetilde{\mathbf{v}}_b^H\mathbf{h}_b\widetilde{\boldsymbol{\theta}}_1^H\mathbf{G}_{b1}\mathbf{H}_{s1}\widetilde{\mathbf{v}}_b\},\\
			&c=|\mathbf{h}_b^H\widetilde{\mathbf{v}}_b|^2,~d=|\widetilde{\boldsymbol{\theta}}_2^H\mathbf{G}_{b2}\mathbf{H}_{s2}\widetilde{\mathbf{v}}_e|^2,\\
			&e=\sigma_1^2\|\widetilde{\boldsymbol{\theta}}_1^H\mathbf{G}_{b1}\|^2,~f=\sigma_2^2\|\widetilde{\boldsymbol{\theta}}_2^H\mathbf{G}_{b2}\|^2,\\
			&\hat{a}=\|\widetilde{\boldsymbol{\theta}}_1^H\mathbf{G}_{e1}\mathbf{H}_{s1}\widetilde{\mathbf{v}}_b\|^2,~\hat{b}=|\widetilde{\boldsymbol{\theta}}_2^H\mathbf{G}_{e2}\mathbf{H}_{s2}\widetilde{\mathbf{v}}_e|^2,\\
			&\hat{c}=\mathfrak{R} \{\widetilde{\mathbf{v}}_e^H\mathbf{h}_e\widetilde{\boldsymbol{\theta}}_2^H\mathbf{G}_{e2}\mathbf{H}_{s2}\mathbf{v}_e\},~\hat{d}=|\mathbf{h}_e^H\widetilde{\mathbf{v}}_e\|^2,\\
			&\hat{e}=\sigma_1^2\|\widetilde{\boldsymbol{\theta}}_1^H\mathbf{G}_{e1}\|^2,~\hat{f}=\sigma_2^2\|\widetilde{\boldsymbol{\theta}}_2^H\mathbf{G}_{e2}\|^2,\\
			&A=\eta\beta P_s\|\widetilde{\boldsymbol{\theta}}_1^H\text{diag}(\mathbf{H}_{s1}\widetilde{\mathbf{v}}_b)\|^2+\sigma_1^2,\\
			&B=\eta(1-\beta)P_s\|\widetilde{\boldsymbol{\theta}}_2^H\text{diag}(\mathbf{H}_{s2}\widetilde{\mathbf{v}}_e)\|^2+\sigma_2^2.
		\end{align}
	\end{subequations}
	Then taking (\ref{rou1}) and (\ref{rou2}) into (\ref{gamma_b}) and (\ref{gamma_e}), respectively, the SINRs of Bob and Eve are rewritten as
	\begin{subequations}
		\begin{align}
			&\hat{\gamma}_b(\eta,\beta)=\frac{\gamma_{b1}}{\gamma_{b2}},\\
			&\hat{\gamma}_e(\eta,\beta)=\frac{\gamma_{e1}}{\gamma_{e2}},
		\end{align}
	\end{subequations}
	where
	\begin{subequations}
		\begin{align}
			\gamma_{b1}=&\eta\beta P_s[(a(1-\eta)\mu P_sB+2b B\sqrt{(1-\eta)\mu P_sA}+cAB],\\
			\gamma_{b2}
			=&d\eta(1-\eta)(1-\beta)(1-\mu)P_s^2A+e(1-\eta)\mu P_s B\\
			&+f(1-\eta)(1-\mu) P_s A+\sigma_b^2AB,\nonumber\\
			\gamma_{e1}=&\eta(1-\eta)\beta\mu P_s^2\hat{a}B,\\
			\gamma_{e2}
			=&\eta(1-\beta) P_s[(\hat{b}(1-\eta)(1-\mu) P_sA\\
			&+2\hat{c}A\sqrt{(1-\eta)(1-\mu )P_sB}+\hat{d}AB]+\hat{e}B(1-\eta)\mu P_s\nonumber\\
			&+\hat{f}(1-\eta)(1-\mu)P_sA+\sigma_e^2AB.\nonumber
		\end{align}
	\end{subequations}
	Hence, given $\rho_k$, $\widetilde{\mathbf{v}}_b$, $\widetilde{\mathbf{v}}_e$ and $\widetilde{\boldsymbol{\theta}}_k$, the optimization problem related to $\eta$ and $\beta$ are written as below
	\begin{subequations}
			\begin{align}
			&\max_{\eta,\beta}~~~~~~~	\text{SR}_1(\eta,\beta)=\frac{\log_2(1+\hat{\gamma}_b(\eta,\beta))}{\log_2(1+\hat{\gamma}_e(\eta,\beta))}\label{SR1}\\
			&~~\text{s.t.}~~~~~~~~
			0\le \beta \le 1,~0\le\eta\le 1.
		\end{align}
	\end{subequations}
	To tackle the above problem, three traditional search optimization algorithms: ES, PSO and SA are directly applied to search the solutions to $\eta$, $\beta$ in the domain of (0, 1). 
	
	For PSO algorithm, the feasible solutions to the above problem are denoted as the positions of particles. Here, the search space is defined as a three-dimensional (3D) space and each position of particle is correspondingly denoted as ($\eta$, $\beta$, $\text{SR}_1$). The velocity $\mathbf{q}$ and position $\mathbf{p}$ of $i$-th particle is updated in line with
		\begin{subequations}
		\begin{align}
			&\mathbf{q}_i(t) = w\mathbf{q}_i(t-1) + c_1  \text{rand}_1(2)(\mathbf{m}_{\text{best},i} - \mathbf{p}_i(t-1)) \label{PSO_v} \nonumber\\
			&~~~~~~\quad + c_2\text{rand}_2(2) (\mathbf{n}_{\text{best}} - \mathbf{p}_i(t-1)),\\
			&\mathbf{p}_i(t) = \mathbf{p}_i(t-1) + \mathbf{q}_i(t),\label{PSO_x}
		\end{align}
	\end{subequations}
	where $t$ is the iteration number. Besides, $w$, $c_1$ and $c_2$ represent inertia weight, individual and social cognitive items, respectively. 
	$\mathbf{m}_{\text{best},i}$ and $\mathbf{n}_{\text{best}}$ denote the best position of the $i$-th particle and the best position of the entire particle swarm, respectively. Afterwards, particle swarm collaborate and share their position information, so that the global optimal solution is obtained. Subsequently, the optimal solution to the objective function is obtained.
	
    For SA algorithm, temperature has an important impact on the current energy, which is the value of objective function in optimization problem. Specifically, when the temperature is high, the energy value changes significantly. On the contrary, when the temperature is low, the change in energy value is relatively slow. Firstly, we need to set the cooling rate, initial temperature, $\eta$ and $\beta$. By adjusting the temperature, we can search some new energy values in the adjacent intervals of $\eta$ and $\beta$. Based on the energy difference function (\ref{DelatE}) and the probability density function (\ref{DelatP}), we decide whether to accept this new energy value. By repeating the cooling process, the global optimal solution of the objective function can be obtained.
	\begin{subequations}
		\begin{align}
			& \Delta  = Q(i+1) - Q(i)\label{DelatE},\\
			&P(\Delta , T) = e^{-\frac{\Delta }{T}}, \label{DelatP}
		\end{align}
	\end{subequations}
	where $Q(i)$ and $Q(i+1)$ denote the $i$-th and $i+1$-th energy value in the cooling process, and $\Delta$ is the energy difference. Additionally, $P(\Delta, T)$ represents the probability density function for accepting the new energy value.
	
	In addition, the computational complexity of ES, PSO, SC and SA are $\mathcal{O}\{K_{ES}\}$,
	$\mathcal{O}\{K_{PSO}I_{PSO}\}$ and $\mathcal{O}\{K_{SA}I_{SA}\}$, where $K_{ES}$, $I_{SC}$ and $I_{SA}$ denote the total search points of ES and the iterations of PSO and SA. Besides, $K_{PSO}$ and $K_{SA}$ are the number of particles and the annealing times of SA, respectively. Thus the  computational complexity of NSP-MRR-PA (ES), NSP-MRR-PA (PSO) and NSP-MRR-PA (SA) are $\mathcal{O}\{NM+N+K_{ES}\}$, $\mathcal{O}\{NM+N+K_{PSO}I_{PSO}\}$ and  $\mathcal{O}\{NM+N+K_{SA}I_{SA}\}$.
	\begin{algorithm}
		\caption{ Proposed NSP-MRR-PA Algorithm}
		\begin{algorithmic}[1]
			\STATE Initialize $\widetilde{\mathbf{v}}_b$ and $\widetilde{\mathbf{v}}_e$.
			\STATE Set $i=1$.
			\REPEAT
			\STATE Update $\widetilde{\mathbf{v}}_e(i)$ and $\widetilde{\mathbf{v}}_b(i)$ by (\ref{v_e}) and (\ref{v_b}).
			\STATE Update $\widetilde{\boldsymbol{\theta}}_1(i)$ and $\widetilde{\boldsymbol{\theta}}_2(i)$ by (\ref{MRR_theta_b}) and (\ref{MRR_theta_e}).
			\STATE Update $\rho_1(i)$ and $\rho_2(i)$ by (\ref{rou1}) and (\ref{rou2}).
			\STATE Search $\eta(i)$ and $\beta(i)$ by ES, PSO and SA method.
			\STATE Set $i = i + 1$.
			\UNTIL $\|\widetilde{\mathbf{v}}_b(i+1) - \widetilde{\mathbf{v}}_b(i)\| \le 10^{-4}$ and  $\|\widetilde{\mathbf{v}}_e(i+1) - \widetilde{\mathbf{v}}_e(i)\| \le 10^{-4}$.
			\STATE  Take  $\eta(i)$, $\beta(i)$, $\rho_1(i)$, $\rho_2(i)$, $\widetilde{\mathbf{v}}_b(i)$, $\widetilde{\mathbf{v}}_e(i)$, $\widetilde{\boldsymbol{\theta}}_1(i)$ and $\widetilde{\boldsymbol{\theta}}_2(i)$ into (\ref{SR11}), $\rm {SR}_1$ can be obtained.
		\end{algorithmic}
	\end{algorithm}

	\section{SIMULATION RESULTS AND DISCUSSIONS}
	In this section,  the simulation parameters are set as follows: the position of BS, $1$-th IRS, $2$-th IRS, Bob, and Eve are (0m, 0m, 0m), (80m, 20m, 30m), (80m, 30m, 20m), (100m, 15m, 0m) and (120m, 5m, 0m), respectively. Additionally, the number of BS antennas is set as $M=$ 8. It is assumed that all channels are LOS channel, and the path loss at 1m is set as  $-30$dBm. For the other parameters, we have the settings as follow: $\mu=$ 0.8 and $\sigma_1^2=\sigma_2^2=\sigma_b^2=\sigma_e^2=-70$dBm.
	\begin{figure}[t]
		\centering
		\includegraphics[width=0.5\textwidth]{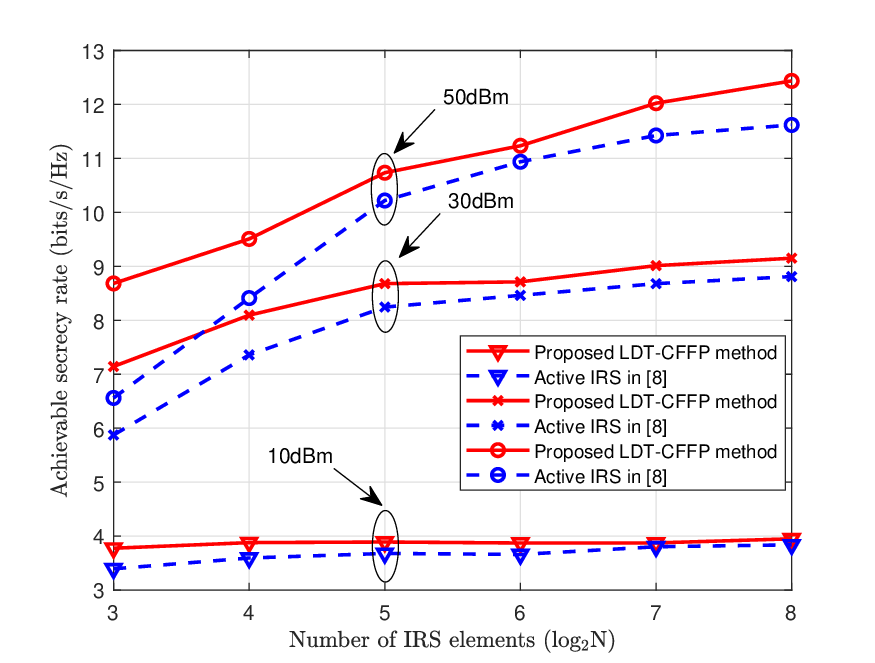}
		\caption{SR versus the number of IRS elements $N$ with three different SNRs}
		\label{Proposed_N}
	\end{figure}
	
	Fig. \ref{Proposed_N} depicts SR curves of LDT-CFFP method with the total power constraint and the method in \cite{DLL9998527} for various total power levels. From Fig. \ref{Proposed_N}, it's evident that LDT-CFFP method attain a higher SR compared to the method in \cite{DLL9998527}. As the total power $P_s$ increases, the SRs of the two methods increase. Meanwhile, the gap between LDT-CFFP and the  method in \cite{DLL9998527} becomes more obvious. Particularly, the maximum gap can reach up to 2 bits when the nuber of IRS elements is 8.
	
	\begin{figure}[t]
		\centering
		\includegraphics[width=0.5\textwidth]{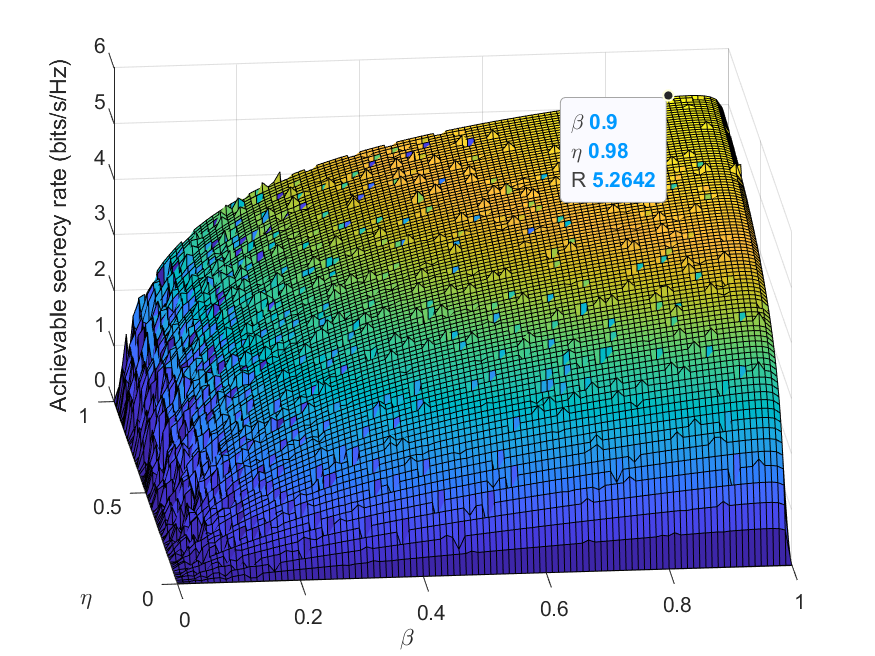}
		\caption{3D SR surface of NSP-MRR-PA (ES) method versus the PA factors $\eta$ and $\beta$ with $N_1=N_2=$ 32}
		\label{ES_1}
	\end{figure}
	
	Fig. \ref{ES_1} illustrates a 3D SR surface of NSP-MRR-ES method versus the PA factors $\eta$ and $\beta$ with $N1=N2=$ 32 and $P_s=$ 20 dBm. It is demonstrated that the PA factors $\eta$ and $\beta$ have significant impact on SR. Obviously, NSP-MRR-PA (ES) method has the global maximum SR when ($\eta, \beta) =$ (0.98, 0.90). 

	\begin{figure}[t]
		\centering
		\includegraphics[width=0.5\textwidth]{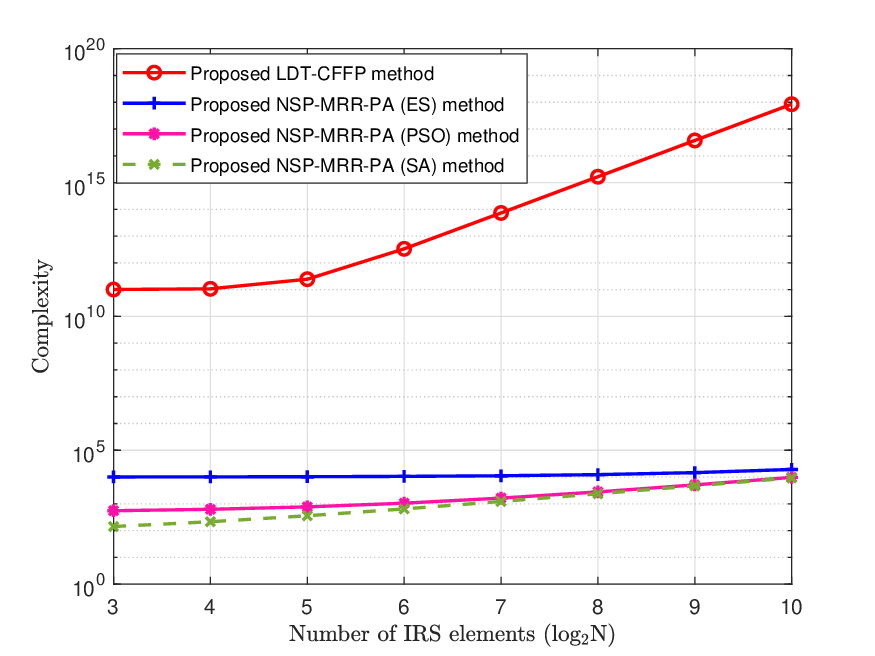}
		\caption{Computational complexity versus the number of IRS elements $N$}
		\label{complexity}
	\end{figure}
	
	Fig. \ref{complexity} plots the complexity curves of LDT-CFFP and  NSP-MRR-PA (ES), NSP-MRR-PA (PSO) and NSP-MRR-PA (SA) algorithms versus the number of IRS elements. From Fig. \ref{complexity}, as the number of IRS elements increases, the computational complexity of the proposed methods gradually increases.
	It's evident that the complexity of LDT-CFFP is far higher than those of NSP-MRR-PA (ES), NSP-MRR-PA (PSO) and NSP-MRR-PA (SA) schemes.

	\begin{figure}[t]
		\centering
		\includegraphics[width=0.5\textwidth]{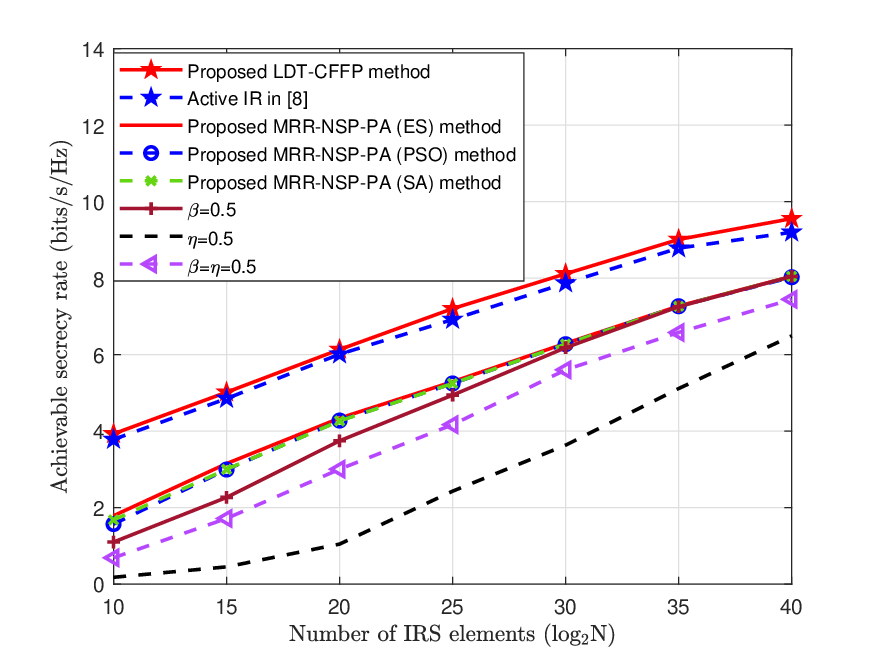}
		\caption{SR versus total power $P$ with $N=$ 32 and $N_1=N_2=$ 16}
		\label{P_s}
	\end{figure}	
	
	Fig. \ref{P_s} shows the SR curves of the proposed LDT-CFFP, NSP-MRR-PA (ES), NSP-MRR-PA (PSO) and NSP-MRR-PA (SA) methods and the other PA strategies with fixed PA factors versus the total power $P$. The SRs obtained by NSP-MRR-PA (ES), NSP-MRR-PA (PSO) and NSP-MRR-PA (SA) methods are higher than those of the PA strategies with fixed PA factors, i.e., $\eta=$ 0.5, $\beta=$ 0.5 and $\eta=\beta=$ 0.5, which proves the effectiveness and superiority of ES, PSO, SC and SA schemes. Moreover, it is shown that LDT-CFFP method can achieve 2.1 bits SR gain over NSP-MRR-PA (ES), NSP-MRR-PA (PSO) and NSP-MRR-PA (SA) methods.

	\begin{figure}[t]
		\centering
		\includegraphics[width=0.5\textwidth]{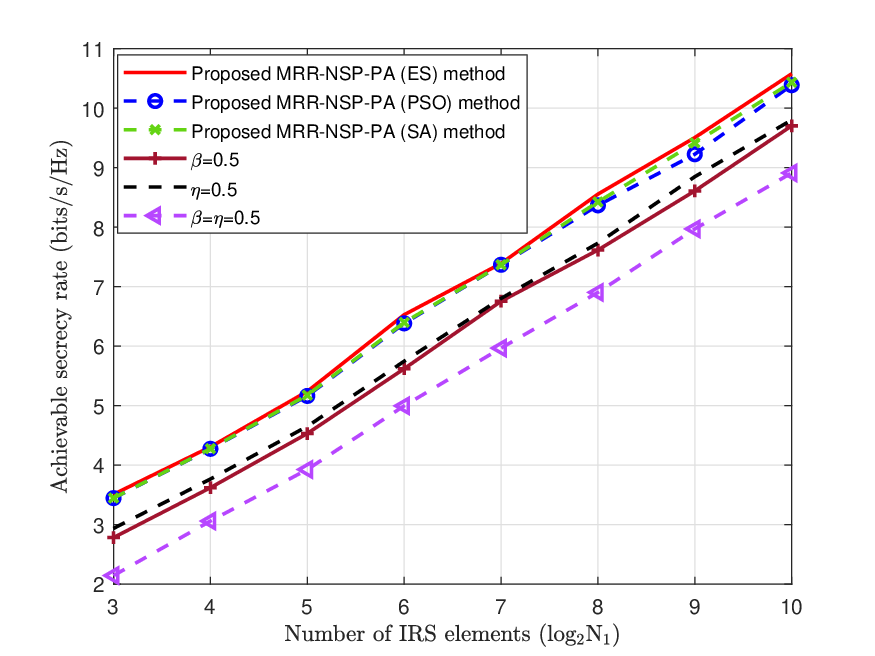}
		\caption{SR versus number of 1-th IRS elements $N_1$ with $N_2=$ 32}
		\label{N1_1}
	\end{figure}
	
\begin{figure}[t]
		\centering
		\includegraphics[width=0.5\textwidth]{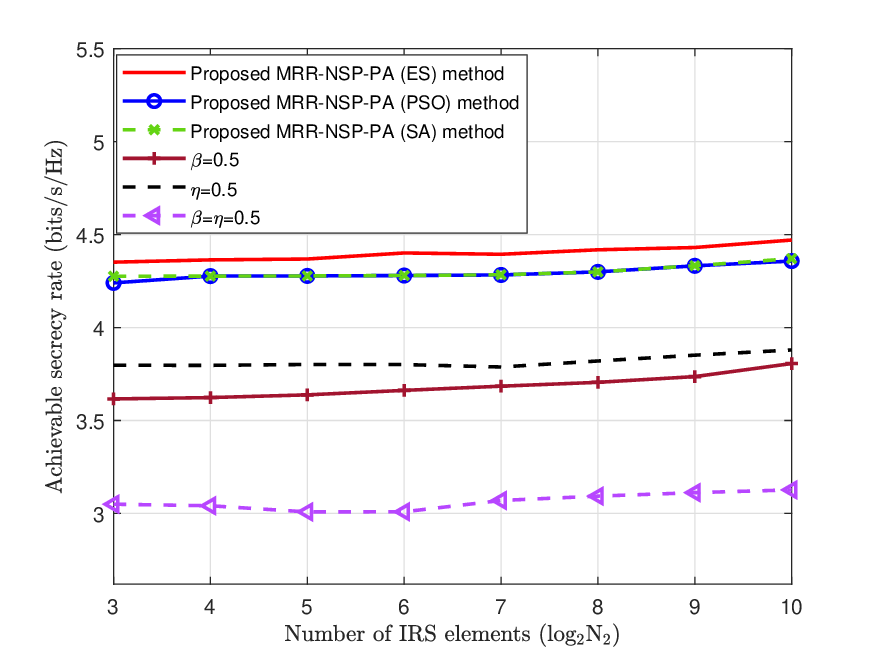}
		\caption{SR versus number of 2-th IRS elements $N_2$ with $N_1$=32}
		\label{N2_1}
	\end{figure}

	Fig. \ref{N1_1} and Fig. \ref{N2_1} plot the SR curves of the proposed NSP-MRR-PA (ES), NSP-MRR-PA (PSO) and NSP-MRR-PA (SA) algorithms versus the numbers of IRS elements of 1-th block and 2-th block, respectively. From Fig. \ref{N1_1}, it is shown that as the number $N_1$ of elements of 1-th IRS block  increases, the SRs of NSP-MRR-PA (ES), NSP-MRR-PA (PSO) and NSP-MRR-PA (SA) algorithms significantly improves. However, from Fig. \ref{N2_1}, it is seen that the SRs of NSP-MRR-PA (ES), NSP-MRR-PA (PSO) and NSP-MRR-PA (SA) schemes increase slowly with the growth of the number $N_2$ of 2-th IRS block units. Besides, in regardless of the increasement of $N_1$ and $N_2$, it is validated that the proposed NSP-MRR-PA (ES), NSP-MRR-PA (PSO) and NSP-MRR-PA (SA) algorithms always perform better than the PA strategies with fixed PA factors, i.e., $\eta=$0.5, $\beta=$0.5 and $\eta=\beta=$0.5.

	\begin{figure}[t]
		\centering
		\includegraphics[width=0.5\textwidth]{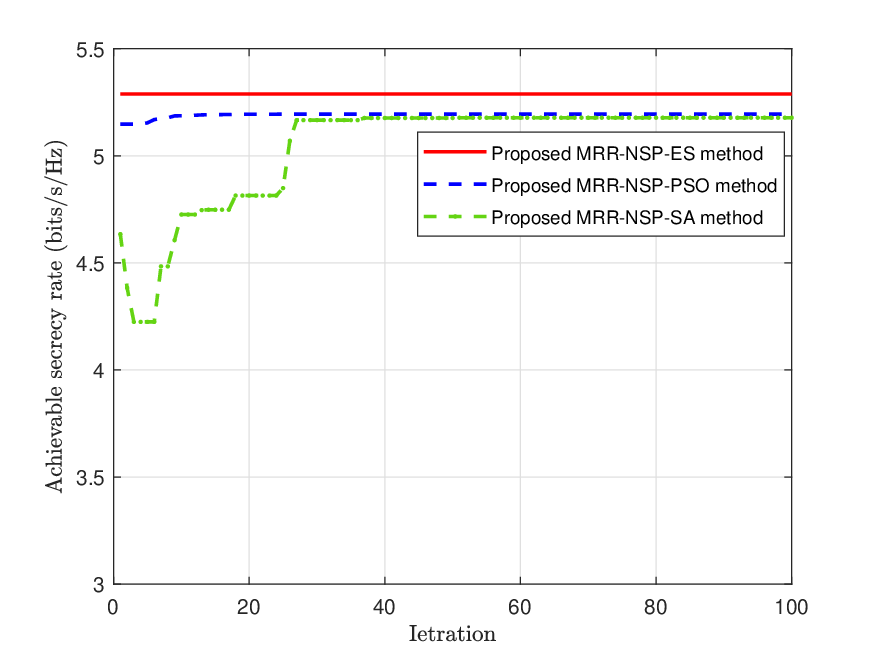}
		\caption{SR versus iterations of ES, PSO and SA algorithms  }
		\label{shoulian}
	\end{figure}

	Fig. \ref{shoulian} verifies the proposed NSP-MRR-PA (ES), NSP-MRR-PA (PSO) and NSP-MRR-PA (SA) methods are convergent after some iterations. In terms of convergence speed, their order is ES$>$PSO$>$SA.

	\section{CONCLUSION}
	
	In this paper, an active IRS-aided DM secure communication system is investigated. To address the SR optimization problem, we first propose a high-performance LDT-CFFP method for solving BS beamforming vectors and IRS reflection coefficient matrix. To reduce the complexity of the LDT-CFFP method, we designed a blocked IRS-based DM secure network and propose NSP-MRR to obtain CF solutions to BS beamforming vectors and IRS reflection coefficient matrix. Moreover, PA strategies were respectively put forward between BS and IRS, CM and AN. ES, PSO and SA methods were adopted to search PA factors. Simulation results displayed that ES, PSO and SA methods achieved better SR than some PA strategies with fixed PA factors, i.e.,  $\eta=0.5$, $\beta=0.5$ and $\eta=\beta=0.5$. Additionally, LDT-CFFP method was superior to NSP-MRR-PA (ES), NSP-MRR-PA (PSO) and NSP-MRR-PA (SA) methods.


\begin{thebibliography}{00}
		\bibitem{TY9530396} F. Shu, Y. Teng, J. Li, M. Huang, W. Shi, J. Li, Y. Wu, and J. Wang,''Enhanced Secrecy Rate Maximization for Directional Modulation Networks via IRS,'' \textit{IEEE Trans. Commun}., vol. 69, no. 12, pp. 8388–8401, Dec. 2021.
		\bibitem{WSM8315448} S. Wan, F. Shu, J. Lu, G. Gui, J. Wang, G. Xia, Y. Zhang, J. Li, and J. Wang, ''Power Allocation Strategy of Maximizing Secrecy Rate for Secure Directional Modulation Networks,'' \textit{IEEE Access}, vol. 6, pp. 38 794–38 801, 2018.
		\bibitem{ST8839973} F. Shu, T. Shen, L. Xu, Y. Qin, S. Wan, S. Jin, X. You, and J. Wang, ''Directional modulation: A physical-layer security solution to B5G and Future Wireless Networks,'' \textit{IEEE Net.}, vol. 34, no. 2, pp. 210–216, Mar. 2020.
		\bibitem{TYarc} Y. Teng, J. Li, M. Huang, L. Liu, G. Xia, X. Zhou, F. Shu, and J. Wang, ''Low-complexity and High-performance receive Beamforming for Secure Directional Modulation Networks against an Eavesdropping-enabled Full-Duplex Attacker,'' \textit{SCIENCE CHINA Inform. Sci.}, vol. 65, no. 1, p. 119302, Dec. 2022.
		\bibitem{QYL8290952} F. Shu, Y. Qin, T. Liu, L. Gui, Y. Zhang, J. Li, and Z. Han, ''Low-Complexity and High-Resolution DOA Estimation for Hybrid Analog and Digital Massive MIMO Receive Array,'' \textit{IEEE Trans. Commun.}, vol. 66, no. 6, pp. 2487–2501, Jun. 2018.
		\bibitem{QJP9120206} J. Qiao and M.-S. Alouini, ''Secure Transmission for Intelligent Reflecting Surface-Assisted mmWave and Terahertz Systems,'' \textit{IEEE Wire. Commun. Let.}, vol. 9, no. 10, pp. 1743–1747, Oct. 2020.
		\bibitem{EMA9086766} M. A. ElMossallamy, H. Zhang, L. Song, K. G. Seddik, Z. Han, and G. Y. Li, ''Reconfigurable Intelligent Surfaces for Wireless Communications: Principles, Challenges, and Opportunities,'' \textit{IEEE Trans. Cog. Commun. Net.}, vol. 6, no. 3, pp. 990–1002, Sep. 2020.
		\bibitem{DLL9998527} Z. Zhang, L. Dai, X. Chen, C. Liu, F. Yang, R. Schober, and H. V. Poor, ''Active RIS vs. Passive RIS: Which will prevail in 6G?'' \textit{IEEE Trans. Commun.}, vol. 71, no. 3, pp. 1707–1725, Mar. 2023.
		\bibitem{CM8723525} M. Cui, G. Zhang, and R. Zhang, ''Secure Wireless Communication via Intelligent Reflecting Surface,'' \textit{IEEE Wire. Commun. Let.}, vol. 8, no. 5, pp. 1410–1414, Oct 2019.
		\bibitem{YYH10284897} Y. Yang, S. Shen, Y. She, W. Wang, B. Yang, and Y. Gao, ''Joint Covert and Secure Communications for Intelligent Reflecting Surface (IRS)-Aided Wireless Networks,'' \textit{in 2023 International Conference on Networking and Network Applications (NaNA)}, Aug. 2023, pp. 138–142.
		\bibitem{LRQ9679804} R. Liu, Q. Wu, M. Di Renzo, and Y. Yuan, ''A Path to Smart Radio
		Environments: An Industrial Viewpoint on Reconfigurable Intelligent	Surfaces,'' \textit{IEEE Wire. Commun.}, vol. 29, no. 1, pp. 202–208, 2022.
		\bibitem{WXH9687669} X. Wang, F. Shu, W. Shi, X. Liang, R. Dong, J. Li, and J. Wang, ''Beamforming Design for IRS-Aided Decode-and-Forward Relay Wireless Network,'' \textit{IEEE Trans. Green Commun. and Net.}, vol. 6, no. 1, pp. 198–207, Mar. 2022.
		\bibitem{LXH10190735} X. Li, C. You, Z. Kang, Y. Zhang, and B. Zheng, ''Double-Active-IRS Aided Wireless Communication with Total Amplification Power Constraint,'' \textit{IEEE Commun. Let.}, vol. 27, no. 10, pp. 2817–2821, Oct. 2023.
		\bibitem{LRQ10298069} R. Liu, H. Lin, H. Lee, F. Chaves, H. Lim, and J. Skld, ''Beginning
		of the Journey Toward 6G: Vision and Framework,'' \textit{IEEE Commun. Magazine}, vol. 61, no. 10, pp. 8–9, 2023.
		\bibitem{FL10036440} L. Ferdouse, I. Woungang, A. Anpalagan, and K. Yamamoto, ''A Resource Allocation Policy for Downlink Communication in Distributed IRS Aided Multiple-Input Single-Output Systems,'' \textit{IEEE Trans. Commun.}, vol. 71, no. 4, pp. 2410–2424, Apr. 2023.
		\bibitem{HM9740154} M. Hua, Q. Wu, and H. V. Poor, ''Power-Efficient Passive Beamforming and Resource Allocation for IRS-Aided WPCNs,'' \textit{IEEE Trans. Commun.}, vol. 70, no. 5, pp. 3250–3265, May 2022.
		\bibitem{LRQ9955484} R. Liu, J. Dou, P. Li, J. Wu, and Y. Cui, ''Simulation and Field Trial
		Results of Reconfigurable Intelligent Surfaces in 5G Networks,'' \textit{IEEE Access}, vol. 10, pp. 122 786–122 795, 2022.
		\bibitem{PCH9201173} S. Hong, C. Pan, H. Ren, K. Wang, and A. Nallanathan, ''Artificial-Noise-Aided Secure MIMO Wireless Communications via Intelligent Reflecting Surface,'' \textit{IEEE Trans. Commun.}, vol. 68, no. 12, pp. 7851–7866, Dec. 2020.
		\bibitem{YRL10288199} R. Ye, Y. Peng, F. Al-Hazemi, and R. Boutaba, ''A Robust Cooperative Jamming Scheme for Secure UAV Communication via Intelligent Reflecting Surface,'' \textit{IEEE Trans. Commun.}, vol. 72, no. 2, pp. 1005–1019, Feb. 2024.
		\bibitem{NHH9234391} N. Hehao and L. Ni, ''Intelligent Reflect Surface Aided Secure Transmission in MIMO Channel with SWIPT,'' \textit{IEEE Access}, vol. 8, pp. 192 132–192 140, 2020.
		\bibitem{AH9399840} H. Albinsaid, K. Singh, A. Bansal, S. Biswas, C.-P. Li, and Z. J. Haas, ''Multiple Antenna Selection and Successive Signal Detection for SM-Based IRS-Aided Communication,'' \textit{IEEE Signal Processing Let.}, vol. 28, pp. 813–817, 2021.
		\bibitem{LLL9232092} L. Lai, J. Hu, Y. Chen, H. Zheng, and N. Yang, ''Directional Modulationenabled Secure Transmission with Intelligent Reflecting Surface,'' \textit{in 2020 IEEE 3rd International Conference on Information Communication and Signal Processing (ICICSP)}, Sep. 2020, pp. 450–453.
		\bibitem{LYQ10172219} Y. Lin, B. Shi, F. Shu, R. Dong, P. Zhang, and J. Wang, ''Enhanced Secure Wireless Transmission Using IRS-Aided Directional Modulation,'' \textit{IEEE Trans. Veh. Technol.}, vol. 72, no. 12, pp. 16 794–16 798, Dec. 2023.
		\bibitem{YG9658554} Y. Ge and J. Fan, “Robust Secure Beamforming for Intelligent Reflecting Surface Assisted Full-Duplex MISO Systems,” \textit{IEEE Trans. Inform. Forensics and Security}, vol. 17, pp. 253–264, 2022.
		\bibitem{WXM8333706} F. Shu, X. Wu, J. Hu, J. Li, R. Chen, and J. Wang, ''Secure and Precise Wireless Transmission for Random-Subcarrier-Selection-Based Directional Modulation Transmit Antenna Array,'' \textit{IEEE Jou. Sel. Areas Commun.}, vol. 36, no. 4, pp. 890–904, Apr. 2018.
		\bibitem{LWG9963962} W. Lv, J. Bai, Q. Yan, and H. M. Wang, ''RIS-Assisted Green Secure Communications: Active RIS or Passive RIS?'' \textit{IEEE Wire. Commun. Let.}, vol. 12, no. 2, pp. 237–241, Feb. 2023.
		\bibitem{KZY10235893} Z. Kang, C. You, and R. Zhang, ''Active-Passive IRS Aided Wireless Communication: New Hybrid Architecture and Elements Allocation Optimization,'' \textit{IEEE Trans. Wire. Commun.}, vol. 23, no. 4, pp. 3450–3464, Apr. 2024.
		\bibitem{LJ10135161} F. Shu, J. Liu, Y. Lin, Y. Liu, Z. Chen, X. Wang, R. Dong, and J. Wang, ''Three High-Rate Beamforming Methods for Active IRS-AidedWireless Network,'' \textit{IEEE Trans. Veh. Technol.}, pp. 1–5, 2023.
		\bibitem{LYQ10141981} Y. Lin, F. Shu, R. Dong, R. Chen, S. Feng, W. Shi, J. Liu, and J. Wang, ''Enhanced-Rate Iterative Beamformers for Active IRS-AssistedWireless Communications,'' \textit{IEEE Wire. Commun. Lett.}, vol. 12, no. 9, pp. 1538–1542, Sep. 2023.
		\bibitem{CYW793310} C. Y. Wong, R. Cheng, K. Lataief, and R. Murch, ''Multiuser OFDM with adaptive subcarrier, bit, and power allocation,'' \textit{IEEE Jou. Sel. Areas Commun.}, vol. 17, no. 10, pp. 1747–1758, Oct. 1999.
		\bibitem{ZJH10284767} J. Zhao, X. Guan, and X. Li, ''Power Allocation Based on Genetic Simulated Annealing Algorithm in Cognitive Radio Networks,'' \textit{Chinese Jou. Electr.}, vol. 22, no. 1, pp. 1–4, Jan. 2013.
		\bibitem{KZ8661536} Z. Kaleem, N. N. Qadri, T. Q. Duong, and G. K. Karagiannidis, ''Energy-Efficient Device Discovery in D2D Cellular Networks for Public Safety Scenario,'' \textit{IEEE Sys. Jou.}, vol. 13, no. 3, pp. 2716–2719, Sep. 2019.
		\bibitem{ZXK8051264} X. Zhang, H. Bie, Q. Ye, C. Lei, and X. Tang, “Dual-Mode Index Modulation Aided OFDM with Constellation Power Allocation and Low-Complexity Detector Design,” IEEE Access, vol. 5, pp. 23 871–23 880, 2017.
		\bibitem{NA9780538733519} R. L. B. Richard L Burden, J Douglas Faires, ''Numerical analysis,'' Aug. 2010.
		\bibitem{HL10361533} H. Ibraiwish, M. W. Eltokhey, and M.-S. Alouini, ''Energy Efficient Deployment of VLC-Enabled UAV Using Particle Swarm Optimization,'' \textit{IEEE Open Jou. Commun. Society.}, vol. 5, pp. 553–565, 2024.
		
		
	\end{thebibliography}
\end{document}